%% file: main.tex
\documentclass[sigconf, screen, 10pt]{acmart}

\renewcommand\footnotetextcopyrightpermission[1]{} 

\usepackage{endnotes}
\usepackage{graphicx} 
\usepackage{booktabs} 
\usepackage{xspace} 
\usepackage{balance} 
\usepackage{subfigure} 
\usepackage{xcolor} 
\usepackage{enumitem} 
\usepackage{amsmath} 
\usepackage{comment}

\setcopyright{rightsretained}

\acmDOI{10.475/123_4}

\acmISBN{123-4567-24-567/08/06}

\acmConference[SIGMOD'19]{ACM SIGMOD International Conference on Management of Data}{June 2018}{Amsterdam, The Netherlands} 
\acmYear{2019}
\copyrightyear{2019}
\acmPrice{15.00}

\definecolor{darkred}{rgb}{0.5,0,0}
\definecolor{darkgreen}{rgb}{0,0.5,0}
\definecolor{darkblue}{rgb}{0,0,0.5}
\hypersetup{colorlinks,linkcolor=darkblue,filecolor=darkgreen,urlcolor=darkred,citecolor=darkblue}

\newtheorem{exmp}{Example}

\newcommand{\wts}{\texttt{wts}\xspace}
\newcommand{\rts}{\texttt{rts}\xspace}
\newcommand{\cts}{\texttt{cts}\xspace}
\newcommand{\crts}{\texttt{crts}\xspace}

\newcommand{\codename}{SCAR\xspace}
\newcommand{\twopl}{S2PL\xspace}

\newcommand{\occ}{OCC\xspace}
\newcommand{\rc}{RC\xspace}

\begin{document}

\title{\codename: Strong Consistency using Asynchronous Replication with Minimal Coordination}

\author{Yi Lu}
\affiliation{%
  \institution{MIT CSAIL}
}
\email{yilu@csail.mit.edu}

\author{Xiangyao Yu}
\affiliation{%
  \institution{MIT CSAIL}
}
\email{yxy@csail.mit.edu}

\author{Samuel Madden}
\affiliation{%
  \institution{MIT CSAIL}
}
\email{madden@csail.mit.edu}

\input{abstract}
\settopmatter{printfolios=true} 
\maketitle

\input{introduction}
\input{background}

\input{scar}
\input{ft}

\input{isolation}
\input{optimizations}
\input{discussion}
\input{evaluation}

\input{related_work}

\input{conclusion}

\balance

{\footnotesize \bibliographystyle{ACM-Reference-Format}
\bibliography{references}}

\end{document}

%% file: abstract.tex
\begin{abstract}

Data replication is crucial in modern distributed systems as a means to provide high availability.
Many techniques have been proposed to utilize replicas to improve a system's performance, 
often requiring expensive coordination or sacrificing consistency. 
In this paper, we present \codename, a new distributed and replicated in-memory database that allows serializable transactions to read from backup replicas with minimal coordination. 
\codename works by assigning logical timestamps to database records so that a transaction can safely read from a backup replica without coordinating with the primary replica, because the records cannot be changed up to a certain logical time. 
In addition, we propose two optimization techniques, timestamp synchronization and parallel locking and validation, to
further reduce coordination.
We show that \codename outperforms systems with conventional concurrency control algorithms 
and replication strategies by up to a factor of 2 on three popular benchmarks. 
We also demonstrate that \codename achieves higher throughput by running under reduced isolation levels 
and detects concurrency anomalies in real time. 

\end{abstract}

%% file: introduction.tex
\section{Introduction} \label{sec:introduction}

High availability (HA) is crucial in modern data-oriented applications. 
In clusters with hundreds to thousands of 
servers, failure is a norm rather than an exception. When a failure 
happens, a highly available system is able to mask the failure using
standby servers.  In most applications, high availability is implemented using
 data replication. 

A desirable property of any approach to high availability is strong consistency between replicas, 
i.e., that there is no way for clients to tell when a failover happened, because the state reflected by
the replicas is identical.
Enforcing strong consistency in a replicated and distributed database is a 
challenging task. 
The most common approach is based 
on primary-backup replication, where all reads and writes are handled at the primary replica, which synchronously ships writes to 
the backup replicas.  As a result, the primary releases locks and commits only after writes have propagated to all replicas, blocking other transactions from accessing modified records and 
limiting performance.  In typical configurations, reads are always executed at the primary replica to ensure that a transaction observes the latest data, but this also incurs long  latency if the primary replica is far away from the client and further loads the primary.

If a system could process reads at replicas and asynchro\-nously ship writes to replicas, it could achieve considerably lower latency and higher throughput, because clients can read from the nearest replica, and transactions can release locks before replicas respond to writes.  Indeed, these features are central to many recent systems that offer eventual consistency(e.g., Dynamo~\cite{DeCandiaHJKLPSVV07}).
Observe that both local reads and asynchronous writes introduce the same problem:  the possibility of stale reads at replicas.
Thus, they both introduce the same consistency challenge: the database cannot determine whether the records a transaction reads are consistent or not.
One naive way of consistently reading from backup replicas is to always send a validation request to the primary replica after reading a record (or set of records) at a backup replica, to verify that the records at both replicas are the same. 
However, this also incurs significant network traffic and latency, and requires the primary to be involved in most or all reads, 
and thus is not likely to be better than the traditional primary-backup scheme.
In this paper, we propose an alternative method. In our approach, the primary provides a {\it promise} to a backup replica that a record will not change at the primary for a certain period of {\it logical time}.  
In this way, a transaction can read the backup record without validation at the primary, which significantly reduces the amount of coordination required in transaction processing.

The system we built that embodies this idea is called \textbf{\codename}.  
It is a single-version distributed and replicated data\-base that supports strong consistency, i.e., serializability and snapshot isolation. 
To achieve this goal, \codename implements a logical timestamp-based optimistic concurrency control (OCC) algorithm with critical performance optimizations. 
To our best knowledge, we are the first to use logical timestamps to allow a transaction to read from any backup replica and asynchronously replicate its writes in a replicated data\-base system.

By default, transactions in \codename are serializable. 
In practice, database transactions often execute under reduced isolation levels (e.g., snapshot isolation) for better performance. 
With minor changes to the commit protocol, \codename supports snapshot isolation as well. 
It achieves this without maintaining multiple data versions in the database, and thus requires a smaller storage footprint compared to multi-version concurrency control (MVCC) algorithms. 
Furthermore, \codename provides a low-overhead \textit{concurrency anomaly detector} to report whether each individual transaction running under snapshot isolation actually committed under serializability. 
This allows a user to detect when isolation violations are absent and determine whether the current isolation level of transactions meets the needs of the workload. 

Our evaluation on an eight-server cluster shows that \codename outperforms systems that use conventional concurrency control algorithms and replication strategies by up to factor of 2 on Retwis (a commonly used benchmark that emulates Twitter-like social network), YCSB, and TPC-C.
The performance advantage of \codename is even more significant in the 
Wide-Area Network (WAN) setting. 
Further, \codename is able to achieve 2x performance improvement versus serializability when running under snapshot isolation.
Finally, we show how the concurrency anomaly detector gives a breakdown of concurrency anomalies in real time. 

In summary, this paper makes the following major contributions:

\begin{itemize}[noitemsep,nolistsep]
	\item We present \codename, a distributed and replicated in-memory database. It allows serializable transactions to read from backup replicas and replicates writes asynchronously, with minimal coordination across nodes.
	\item We show how \codename supports snapshot isolation with minor changes to the commit protocol. 
	\item We introduce a concurrency anomaly detection scheme which detects anomaly-free executions in real time with low overhead. 
	\item We propose two optimizations to reduce the overhead of coordination during transaction validation in \codename.
\end{itemize}

%% file: background.tex
\section{Background} \label{sec:background}

This section discusses the background of distributed concurrency control
and data replication.

\subsection{Distributed Concurrency Control}

Concurrency control enforces two critical properties of a database:
atomicity and isolation. Atomicity requires a transaction to expose
either all or none of its changes to the database. The isolation level
specifies when a transaction is allowed to see another transaction's
writes.

Both \textit{serializability} and \textit{snapshot isolation} are considered as strong isolation levels in a distributed database. 
Serializablity (SR) requires transactions to produce the same results as if they are
sequentially executed; it is the gold standard isolation level due to
its robustness and ease of understanding. 
Snapshot isolation (SI) requires a transaction to read a consistent snapshot and perform all
the writes at the same time; but reads can happen earlier than the writes.
Snapshot isolation is weaker than serializability and thus allows more transactions to commit. 
This leads to better performance at the cost of potential concurrency anomalies.

Three classes of concurrency control protocols are commonly used in distributed systems: two-phase 
locking (2PL)~\cite{BernsteinSW79, EswarranGLT76}, optimistic concurrency control (OCC)~\cite{KungR81}, 
and multi-version concurrency control (MVCC)~\cite{Reed78}. 2PL protocols 
are pessimistic and use locks to avoid conflicts. 
An MVCC protocol maintains multiple versions of each tuple in the database. 
This offers higher concurrency since a transaction can potentially pick from amongst several consistent versions to read, at the cost of higher storage overhead and complexity. 
In OCC, a transaction does not acquire locks during execution; after execution, the database validates a transaction to determine whether it commits or aborts.
At low contention, OCC has better performance than 2PL due to its non-blocking execution. 
Although a large number of distributed OCC protocols have been proposed in recent years~\cite{MahmoudANAA14, MuCZLL14, ShuteVSHWROLMECRSA13, YuXPSRD18, ZhangSSKP15}, there has not been a consensus of the best implementation of distributed OCC.
In this paper, we adapt Silo's OCC protocol~\cite{TuZKLM13} to the distributed 
environment and use that to illustrate typical distributed OCC algorithms. 
More details of the baseline will be discussed in Section~\ref{sec:scar}.
Traditionally, all the three classes of concurrency control support serializability, but only MVCC supports snapshot isolation. 
As we show in Section~\ref{sec:isolation}, although \codename is a single-version OCC protocol, it supports both serializability and snapshot isolation without the overhead of storing multiple tuple versions. 

The lifecycle of a distributed transaction contains an execution phase and an atomic commit protocol. 
During the execution phase, a transaction accesses the database and executes transaction logic. 
The atomic commit protocol guarantees that all participating nodes agree on the outcome of the transaction (i.e., commit or abort) and this outcome survives failures. 
In Section~\ref{sec:scar}, we will discuss how a transaction is executed and committed in \codename. 

\subsection{Replication}

Modern database systems support high availability (HA) such that when a subset of servers fail, the rest of the servers can carry out the database functionality, thereby end-users do not notice the server failures. 
High availability requires the database to replicate data across multiple servers and propagate each update to all the replicas. 

Both Paxos-based and primary-backup replication schemes are commonly used in replicated systems. 
Paxos-based~\cite{Lamport01} replication synchronizes each read and write operation to the database. 
While Paxos handles node failures more gracefully, it requires heavy coordination that incurs excessive network traffic and performance degradation~\cite{ZhangSSKP15}.

Primary-backup replication is also a commonly used replication scheme. 
A transaction can write to only the primary copy of each record. 
The primary replica then propagates the write to the backup replicas. 
In a typical primary-backup replication scheme, reads always go to the primary copies in order to observe the latest data. 
However, if the primary copy is on a remote node while a backup copy is local, it is desirable if a transaction can read the local backup copy instead, since it can avoid the network latency. 
Some existing systems have tried to allow transactions to read from backup replicas by having multi-version data storage and global
timestamp allocation~\cite{CorbettDEFFFGGHHHKKLLMMNQRRSSTWW12, PelusoRQ12}. This solution, however, incurs higher storage overhead and complexity for version management; furthermore, generation of consistent timestamps across nodes requires complex protocols. 
Other systems achieve this by only supporting weak consistency levels like causal consistency~\cite{MehdiLCABL17} or eventual consistency~\cite{TerryTPDSH95}. Although this reduces the complexity of managing replication, it introduces concurrency anomalies into transactions, which makes it difficult to implement correct application code for database users. 
In Section~\ref{sec:scar}, we discuss how \codename allows transactions to read from backup copies while enforcing serializability with minimal coordination.

%% file: scar.tex
\section{\codename} \label{sec:scar}

In this section, we first give an overview on \codename to show its benefits over typical distributed OCC protocols. 
We next explain in detail how \codename reads from replicas, validates transactions and applies the writes asynchronously without relaxing the consistency model.
At last, we discuss the non order-preserving serializability that \codename achieves. 

\subsection{Overview} \label{ssec:overview}

The database in \codename is partitioned across a cluster of nodes.
Each partition is mastered on one machine with one or more backup partitions on other nodes.
\codename implements primary-backup replication. 
Each record has a primary replica and one or more backup replicas.
Each replica resides on a different machine. 
If the primary replica of some records fails, one of the backup replicas is promoted as the new primary and the database as a whole does not stop its service. 

\begin{figure}[!t]
    \centering
    \includegraphics[width=0.95\columnwidth]{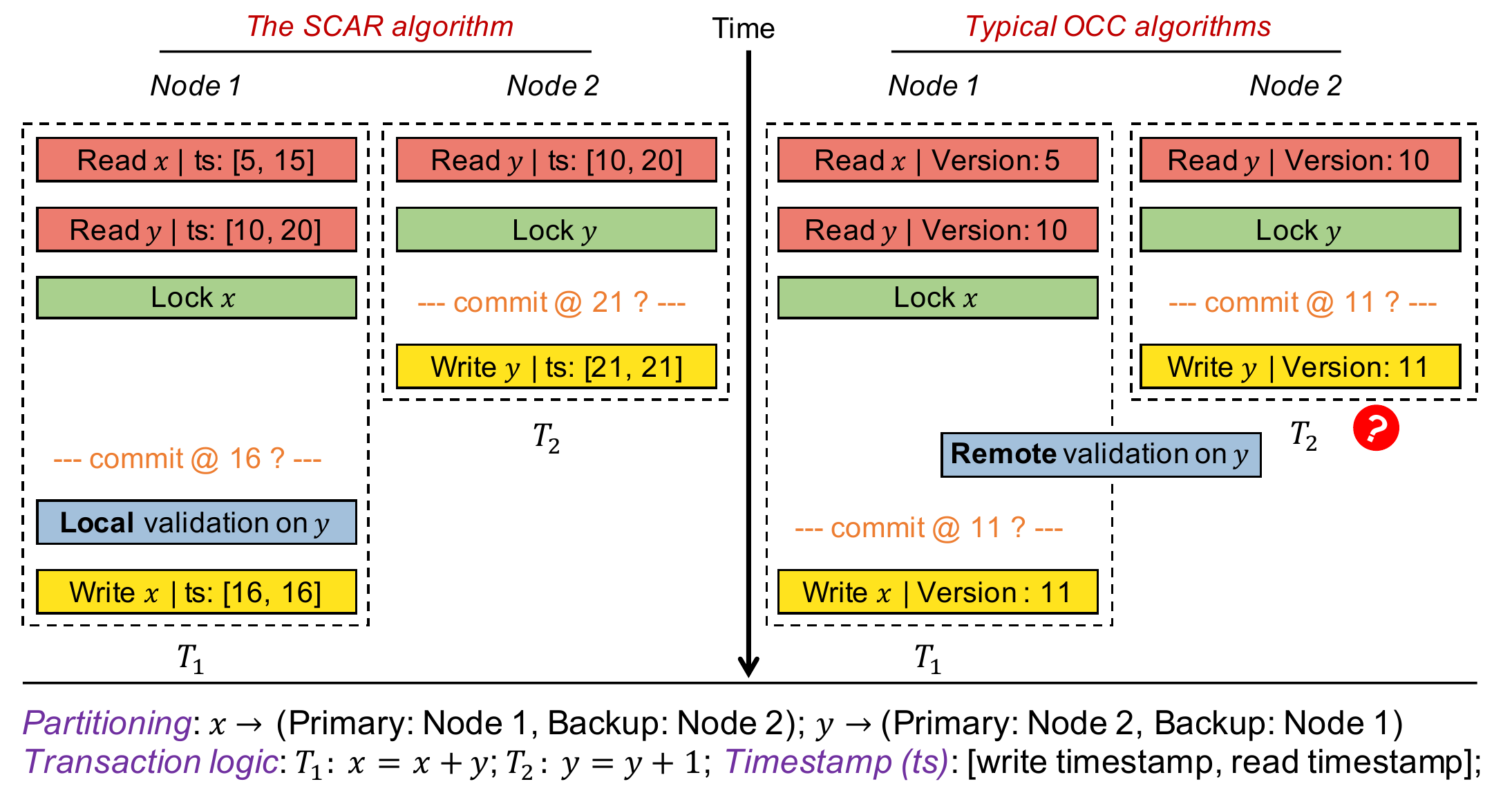}
    \caption{Illustrating the \codename algorithm} \label{fig:scar}
\end{figure}

An important feature of \codename is that a transaction can read from any backup replica without necessarily coordinating with the primary replica. Figure~\ref{fig:scar} illustrates how a pair of transactions (i.e., $T_1$: $x = x + y$; $T_2$: $y = y + 1$) work in \codename and in a distributed OCC system similar to Google's F1~\cite{ShuteVSHWROLMECRSA13}. 
Here, we have a two-node database with two replicas, with two records, $x$ and $y$, each mastered on different nodes.

In a typical distributed OCC protocol (right hand side of the Figure~\ref{fig:scar}), 
the transaction may read from a local replica (\textit{read}($y$) at Node 1), but before the transaction can commit, 
the database must validate the transaction's local read at the primary replica (Node 2), because the database running at Node 1 does not know whether record $y$ has been changed at its primary replica or not.\footnote{The database can avoid this validation if each write locks and updates all the replicas. But this degrades the performance of write operations.}
Such validation introduces round-trip messages which degrade performance.

\codename avoids many of these validations because the data\-base provides per-record \textit{read/write timestamps} to the backup replica as a promise that each record will not be updated until the read timestamp, as shown in the left side of Figure~\ref{fig:scar}. 
Therefore, a transaction does not need to coordinate with the primary replica to validate a record, 
as long as the transaction commits earlier than the record's read timestamp.
In practice, the timestamps can be either physical (e.g., no update can happen in the next 10 seconds) or logical (e.g., no update can happen until the logical timestamp reaches 10). \codename uses the logical timestamp design to avoid some difficulties of physical clocks (e.g., distributed clock synchronization).
Specifically, each record in the database is associated with two logical timestamps, which are represented by two 64-bit integers: \big[$\wts, \rts$\big].
The \wts is the logical write timestamp, indicating when the record was written, and the \rts is the logical {\it read validity} timestamp, indicating that the record can be read at any logical time \textit{ts} such that \wts $\leq \textit{ts} \leq$ \rts.

Suppose the primary nodes of record $x$ and $y$ are Node 1 and Node 2 respectively. 
Similarly, the backup nodes of record $x$ and $y$ are Node 2 and Node 1 respectively.
In the left side of Figure~\ref{fig:scar}, the transaction running on Node 1 reads a local record $x$ which has logical timestamps of \big[5, 15\big]; 
it also reads record $y$ from the local backup replica which has logical timestamps of \big[10, 20\big]. 
The logical read timestamp on record $y$ is a promise that the primary will not update $y$ until at least logical time 21. 
In this example, the transaction can commit locally at Node 1 at timestamp 16 (larger than record $x$'s \rts), 
at which point operations to both $x$ and $y$ are valid; 
and thus there is no need to coordinate with the primary replica of $y$ (i.e., Node 2). 

Logical timestamp-based protocols were first proposed in TicToc~\cite{YuPSD16}.
Unlike TicToc, however, \codename focuses on using logical timestamps to allow transactions to read from backup replicas without relaxing the consistency model and apply writes to the database asynchronously to reduce round-trip communication. 

The rest of this section explains in detail how \codename runs a transaction and manages replication. 
We will discuss how consistency and fault tolerance are achieved in Section~\ref{sec:consistency_and_fault_tolerance}.

\subsection{Reading from Replicas} \label{ssec:reading_from_replicas}

A transaction in SCAR runs in multiple phases: an execution phase, a validation phase, and a commit phase.
We say the node initiating a transaction is the coordinator node, and other nodes are participant nodes.

In the execution phase, a transaction reads records from the database and maintains local copies of them in its \textit{read set} (RS).  
Each entry in the read set contains the value as well as the record's associated \wts and \rts.

For a read request, the coordinator node first checks if the request's primary key is already in the read set. 
This happens when a transaction reads a data record multiple times.
In this case, the coordinator node simply uses the value of the first read.
Otherwise, the coordinator node reads the record from the database.
A record can be read from any replica in \codename. 
To avoid network communication, the coordinator node always reads from its local database of a local copy is available.

\begin{figure}[!t]
\centering
\includegraphics[width=0.95\columnwidth]{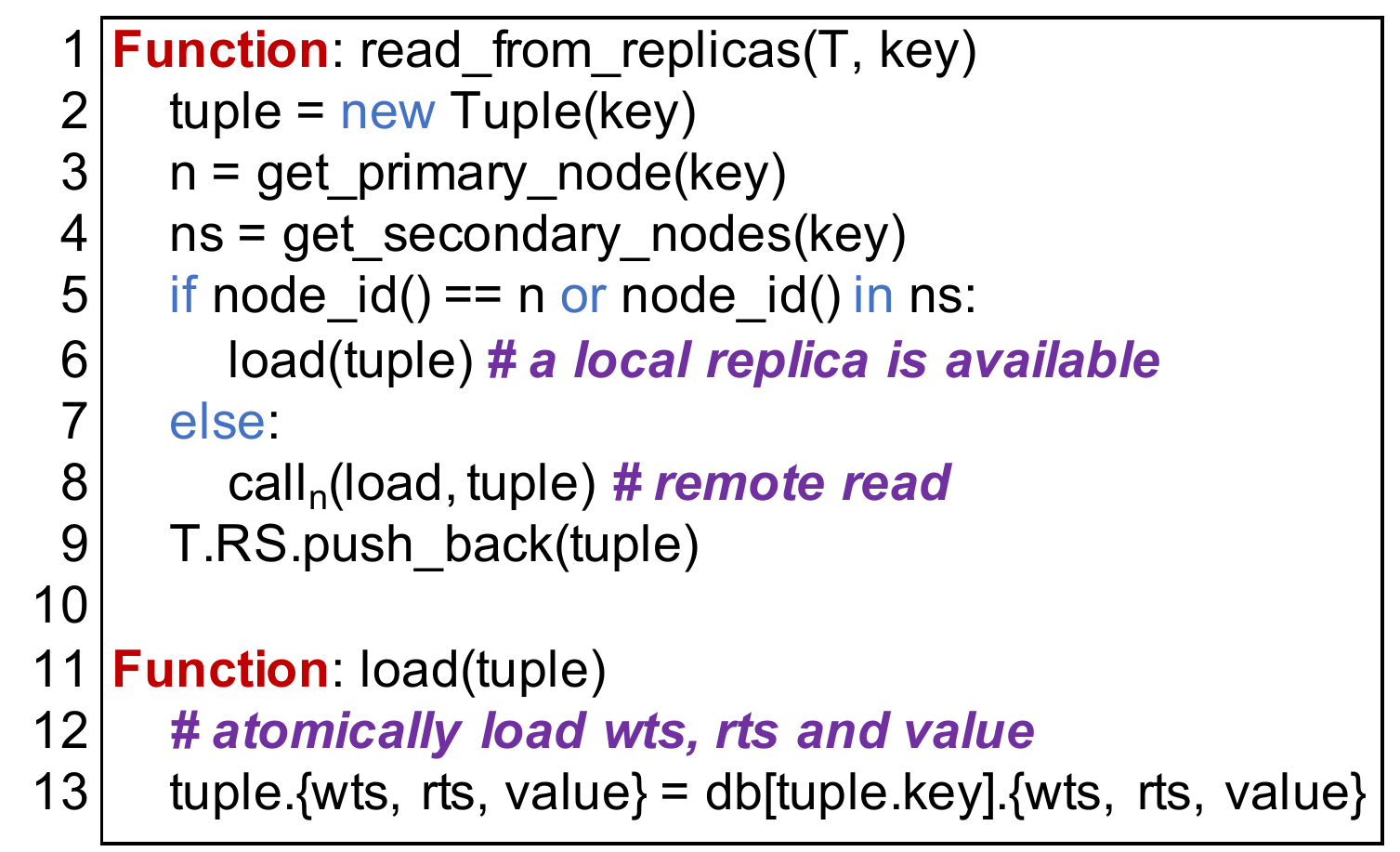}
\caption{Pseudocode to read from replicas} \label{fig:read_from_replicas}
\end{figure}

The coordinator first locates the primary node $n$ and backup nodes $ns$ of the record.
As illustrated in Figure~\ref{fig:read_from_replicas}, a transaction can read the record from its local database in two scenarios: (1) the coordinator node happens to be the primary node $n$. For example, transaction $T_1$ can read record $x$ locally on Node 1 in the left side of Figure~\ref{fig:scar}. 
(2) the coordinator node is a backup node among $ns$, which already has a copy of the record. For example, transaction $T_1$ can read record $y$ locally on Node 1 as well, even though Node 2 is the primary node of record $y$ in the left side of Figure~\ref{fig:scar}.\footnote{Node 1 is a backup node that has a copy of record $y$.}
If no local copy is available, a read request is sent to a participant node, i.e., the remote primary node $n$.

In \codename, logical timestamps (i.e., 64-bit \wts and \rts) are associated with records in both primary and backup replicas. 
For a read request, the system returns both the value and the logical timestamps of a record; and both are stored in the transaction's local read set.
Later in Section~\ref{ssec:transaction_validation}, we explain how logical timestamps are used to avoid coordination during a transaction's validation phase.

All computation is performed in the execution phase. Since \codename's algorithm is optimistic,
 writes are not applied to the database but are stored in a per-transaction \textit{write set} (WS),
in which, as with the read set, each entry has a value and the record's associated \wts and \rts.

For a write operation, if the primary key is not in the write set,
a new entry is created with the value and then inserted into the write set.
Otherwise, the system simply updates the write set with the new value.
Note that for updates to records that are already in the read set, the transaction also copies the \wts and the \rts to the entry in the write set, which are used for validation later on.

\subsection{Transaction Validation} \label{ssec:transaction_validation}

After a transaction finishes its execution phase, it must be successfully validated before it commits. 
We borrow the idea on transaction validation from TicToc~\cite{YuPSD16}, a single-node multi-core concurrency control protocol. 
Different from TicToc, \codename is the first to apply logical timestamps to a distributed and replicated database system to allow a transaction to read from any backup replica and asynchronously replicate its writes using the concept. 
We now describe the three steps to validate a transaction: (1) lock all records in the transaction's write set; (2)  assign a commit timestamp to the transaction; (3) validate all records in the transaction's read set. 
Optimizations involving replication will be introduced later in Section~\ref{sec:optimizations}.

A transaction first tries to acquire locks on each record in the write set to prevent concurrent updates from other transactions.
A locking request is sent to the primary replica of each record. 
To avoid deadlocks, we adopt a \texttt{NO\_WAIT}\footnote{NO\_WAIT dead lock prevention strategy was shown as the most scalable protocol~\cite{HardingAPS17}.} policy, i.e., if the lock is already held on the record,
the transaction does not wait but simply aborts.
For each acquired lock, if any record's latest \wts does not equal to the stored \wts, the transaction aborts as well. 
This is because the record has been changed at the primary replica since the transaction last read it.
The transaction also updates each record's \rts in its local write set in this step.

A commit timestamp \cts is next assigned to the transaction based on all records the transaction accesses (both read set and write set).
The \cts is the smallest timestamp that meets the following two conditions: (1) not less than the \wts of each entry in the read set; (2) larger than the \rts of each entry in the write set. 
To see this, consider the example in the left side of Figure~\ref{fig:scar}, the \cts is 16, which equals to 
\texttt{max}($x$.{\tt rts} + 1, $x$.{\tt wts}, $y$.{\tt wts}).

At last, a transaction validates its read set. 
The transaction's \cts is first compared with the \rts of each record in its read set. 
A read validation request is sent only when a record's \rts is less than the \cts. 
In this case, the transaction tries to extend the record's \rts at the primary node.
The extension would fail in two scenarios: (1) the record's \wts changed, meaning the record was modified by other concurrent transactions; (2) the record is locked by other transactions and the \rts is less than the \cts.
In either case, the \rts cannot be extended and the transaction must abort.
Otherwise, the transaction extends the record's \rts to the transaction's \cts.

\subsection{Asynchronous Write and Replication} \label{ssec:asynchronous_write_and_replication}

\begin{figure}[!t]
\centering
\includegraphics[width=0.95\columnwidth]{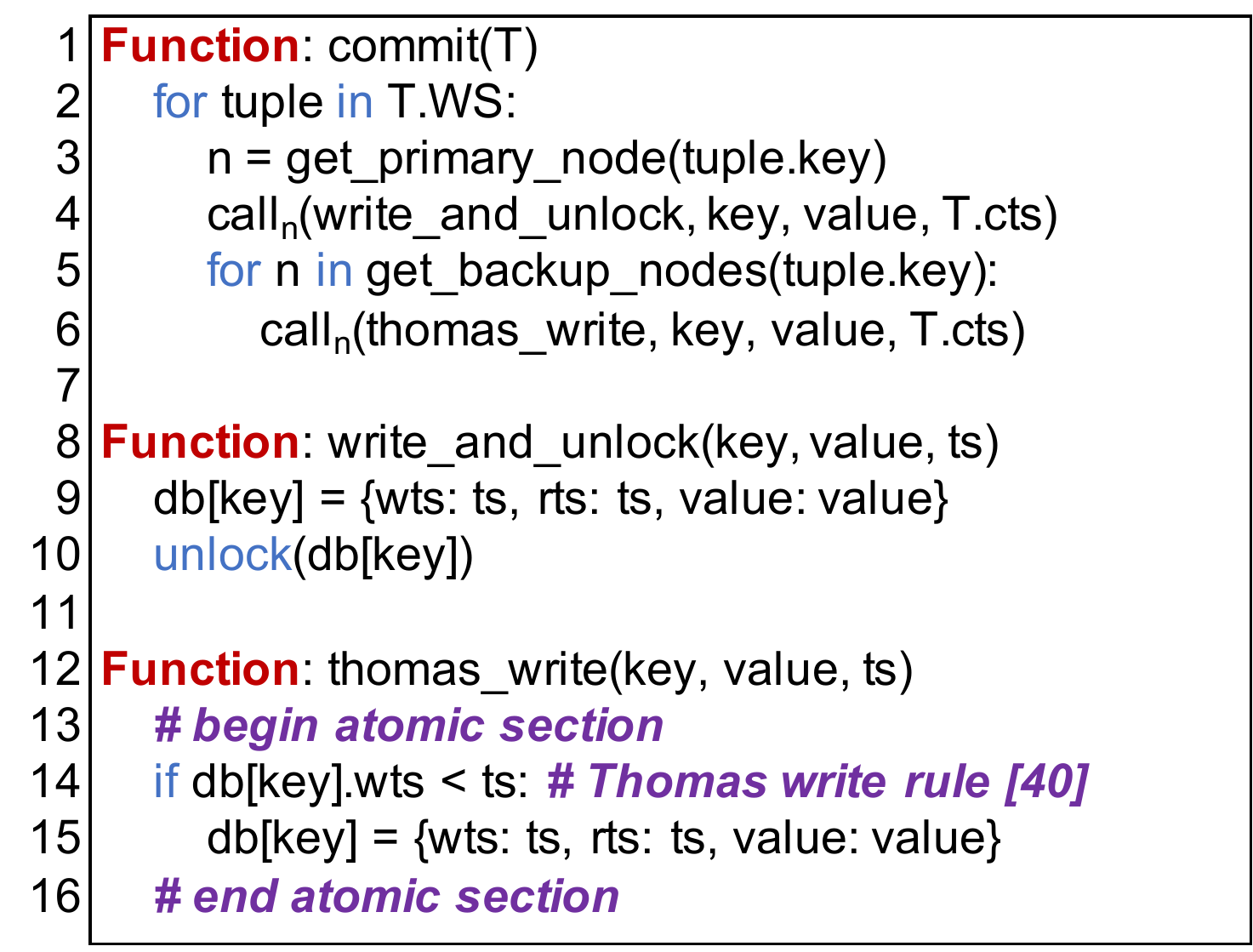}
\caption{Pseudocode to commit a transaction} \label{fig:commit}
\end{figure}

If a transaction fails the validation, it simply aborts, unlocks the acquired locks, and discards its local write set. 
Otherwise, it will commit changes in its write set to the database.  
\codename applies the writes and replication asynchronously to reduce round-trip communication. We will discuss how consistency and fault tolerance are achieved in Section~\ref{sec:consistency_and_fault_tolerance}.

As illustrated in Figure~\ref{fig:commit}, the value of each record in a transaction's write set and the \cts are sent to the primary and backup replicas from the coordinator node by calling the \texttt{commit} function.
There are two scenarios that writes are sent: 
(1) writes are sent to the primary replica: Since the primary replica is holding the lock, 
upon receiving the write request, the primary replica simply updates the value and the logical timestamps for the record in the database to \big[\cts,\cts\big]; 
(2) writes are sent to backup replicas: Since asynchronous replication is employed in \codename, 
upon receiving the write request, the lock on the record is not necessarily held on the primary replica, 
meaning replication requests to the same record from multiple transactions could arrive out of order.
\codename determines whether a replication request at a backup replica should be applied using the Thomas write rule~\cite{Thomas79}: the database applies a write if the \wts of the record in the write request is larger than the current \wts of the record in the database (line 14 -- 15 of Figure~\ref{fig:commit}). Because the \wts of a record monotonically increases in the primary replica, this guarantees that secondary replicas apply the writes in the same order as the order to commit transactions on primary replicas.  

\subsection{Non Order-Preserving Serializability} \label{ssec:non_order_preserving_serializability}

\begin{figure}[!t]
    \centering
    \includegraphics[width=0.99\columnwidth]{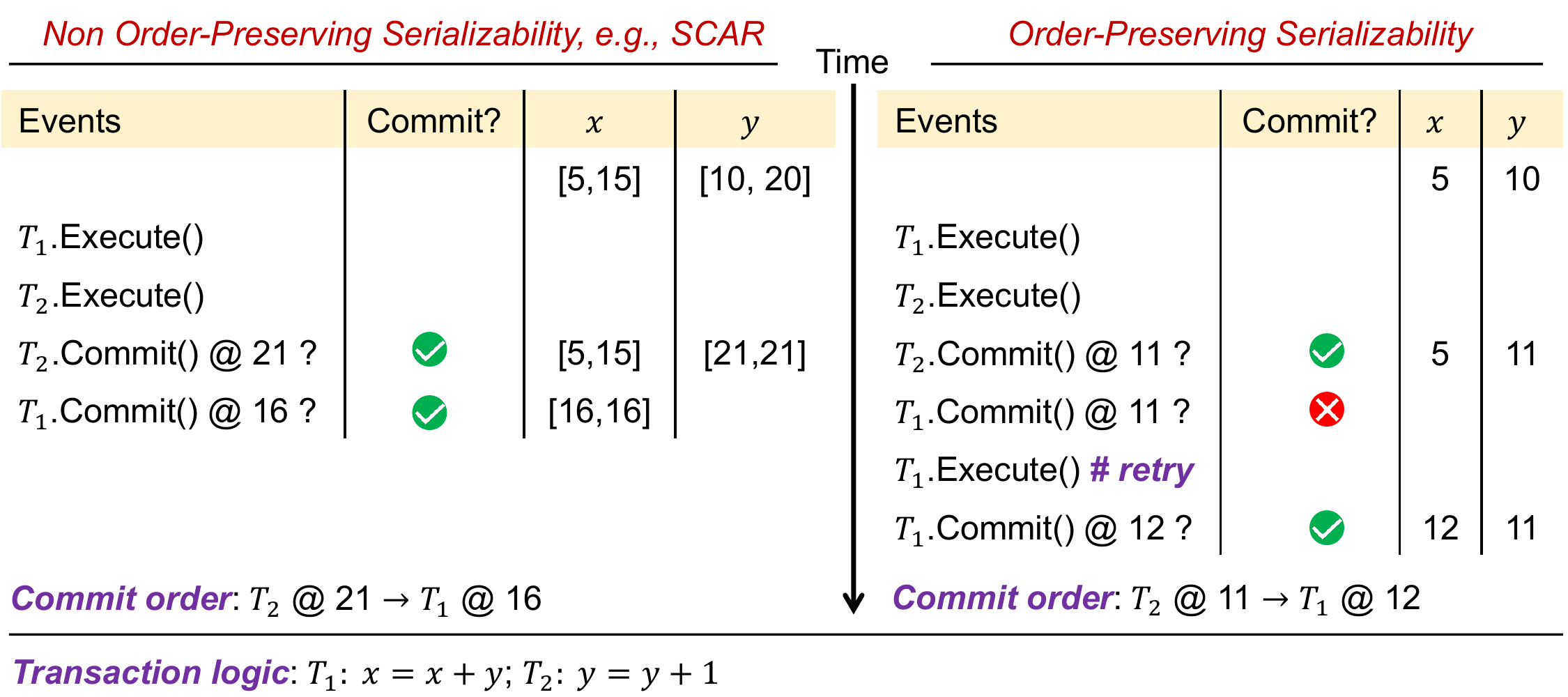}
    \caption{Illustrating non order-preserving serializability vs. order-preserving serializability} \label{fig:order_preserve}
\end{figure}

\codename achieves less coordination in transaction execution but sacrifices external consistency~\cite{CorbettDEFFFGGHHHKKLLMMNQRRSSTWW12}, i.e., the system commits transactions under non order-preserving serializability, which we will describe below. 

Most OCC algorithms are based on physical time (e.g., Silo~\cite{TuZKLM13}). 
In these systems, the database validates a transaction's read set by comparing the data versions from the read set to the latest ones on the primary replicas. 
If any record's primary partition is not on the coordinator, a round-trip communication must be performed when a transaction's read set is validated.
We now use the same example from Figure~\ref{fig:scar} and show the events happening following the physical time in Figure~\ref{fig:order_preserve}. 
Consider the example in the right side of Figure~\ref{fig:order_preserve}, in which transaction $T_1$ and $T_2$ runs concurrently. 
By the time $T_1$ commits, $T_2$ has committed and a new value of record $y$ has been written to the database (i.e., version 11\footnote{The commit time is not less than the version of any record in a transaction's read and write set.}). 
Since record $y$ in $T_1$'s read set has changed from 10 to 11, $T_1$ cannot commit at time 11 and must abort.
$T_1$ must retry and commits at time 12. 
In systems with order-preserving serializability (e.g., Spanner~\cite{CorbettDEFFFGGHHHKKLLMMNQRRSSTWW12}), the commit time of conflicting transactions determines transaction commit order.
  
In contrast, transactions commit in \codename do not necessarily follow the order of commit timestamps, 
i.e., the logical time does not always agree to the physical time. 
Consider the example in the left side of Figure~\ref{fig:order_preserve}.
After transaction $T_2$ commits, which has written a new value of record $y$ (i.e., version \big[21, 21\big]).
Transaction $T_1$ can still commit at time 16, even though record $y$ is in its read set and the value has changed. 
This is because $T_1$'s commit time is earlier than record $y$ last written time in the space of logical time, i.e., 16 falls between logical time 10 and 20. 
Non order-preserving serializability enables \codename to reduce significant network communication when a transaction is validated. 
For example, for each record in a transaction's read set, the system first compares the record's \rts and the transaction's assigned commit timestamp \cts. 
A read validation request is sent only when the record's \rts is less than the \cts and the primary node of the record is not the coordinator node of the transaction. 
Otherwise, the read is already consistent, since it is valid at logical time \cts. 
If all records in the read set can be validated locally, a round trip communication is eliminated entirely.

%% file: ft.tex

\section{Consistency and Fault Tolerance} \label{sec:consistency_and_fault_tolerance}

In this section,  we first describe how \codename ensures consistency with epochs and next show how fault tolerance is achieved. 

\subsection{Ensuring Consistency with Epochs} \label{ssec:ensuring_consistency}

One well known issue with asynchronous replication is the potential for data inconsistency when a failure occurs. In a typical implementation, a transaction commits after successfully updating the primary replica, while the replication requests are still underway. If the primary replica fails after the transaction commits, the replicas are not guaranteed to receive the data. Therefore, the effect of  the last several updates might be lost, leading to inconsistent behavior.

\codename addresses this issue by delaying the commit of a transaction until the completion of its replication as well as the replication of transactions that it depends on. Specifically, \codename borrows ideas from the epoch-based logging scheme used in Silo~\cite{TuZKLM13, ZhengTKL14}, in which transactions commit in batches. A transaction commits only after all transactions within the same batch commit, although a transaction can release its locks early, before the replication process completes. 

Transactions in \codename are separated by epochs (of 10~ms each, by default) using global barriers. Each transaction is assigned the current epoch number as it starts. 
The epoch number advances when the next global barrier is reached. 
Transactions in an epoch commit if all the transactions in this epoch have replicated their write sets to the backup replicas. 
This guarantees that all committed transactions are fully replicated and survive failures. 
It also guarantees that for a committed transaction, all transactions that it depends on have the same or smaller epoch number and thus have committed as well~\cite{ZhengTKL14}.

\subsection{Fault Tolerance} \label{ssec:fault_tolerance}

The system can tolerant up to  $f$ simultaneous failures when each partition has $f + 1$ replicas. For ease of presentation, we discuss the case where only one node fails. For each partition on the failed node, if the primary partition is lost, a secondary partition on other nodes becomes the primary partition.

As discussed above, \codename commits transactions by epochs. Once a fault occurs, \codename rollbacks the database to the last successful epoch, i.e., all tuples that are updated in the current epoch are reverted to the states in the last epoch.
To achieve this, the database maintains two versions of each tuple. One always has the latest value. The other one has the most recent value up to the last successful epoch.  

The system can continue processing transactions when a node fails. Once the failed node restarts, it copies the lost partitions from other nodes. In the meantime, the restarted node uses the Thomas write rule~\cite{Thomas79} to correct its database the same as in Section~\ref{ssec:asynchronous_write_and_replication} and catches up to other nodes using the writes of committed transactions.

%% file: isolation.tex
\section{Isolation Levels} \label{sec:isolation}

\codename supports serializable transactions by default. 
In addition, it also supports snapshot isolation (SI). 
In this section, we first describe the protocol to support SI transactions. 
We then discuss how \codename can be used for monitoring concurrency anomalies from SI transactions in real time. 

\subsection{Transactions under Snapshot Isolation} \label{ssec:transactions_under_snapshot_isolations}

A transaction running under SI does not detect read/write conflicts. 
By not detecting these conflicts, the system is able to achieve a lower abort rate and higher throughput.

Many systems adopt a multi-version concurrency control (MVCC) algorithm to support snapshot isolation. 
In an MVCC-based system, a timestamp is assigned to a transaction when it starts to execute. 
By reading all records that have overlapping time intervals with the timestamp, the transaction is guaranteed to 
observe the state of the database (i.e., a consistent snapshot) at the time when the transaction began.  

Instead of maintaining multiple versions for each record, we made 
minor changes to the algorithm discussed in Section~\ref{sec:scar} to 
support snapshot isolation.
SI transactions do not have to follow a serial order, instead, 
they only require that all reads come from a consistent snapshot of the database and 
there are no conflicts with any concurrent updates made since that snapshot.
\codename achieves this by assigning an additional timestamp to validate the read set of a transaction. 
We introduce a  new timestamp \crts\footnote{\crts is short for {\bf c}ommit {\bf r}ead {\bf t}ime{\bf s}tamp.}, which is the maximum value of the \wts of all records in a transaction's read set. 
The system next uses the \crts to validate the transaction's read set as discussed in Section~\ref{ssec:transaction_validation} 
and ensures all reads are from the state of the database at logical time \crts. 
The system then applies the writes at logical time \cts as before to make sure there are no conflicts with updates.
The \crts is often smaller than the \cts\footnote{The \crts does not have to be larger than the \rts of each entry in the write set.}, which makes a transaction more likely to be validated.
Note that \codename can support a mix of transactions concurrently running under different isolation levels (SI/serializability) as well.

\subsection{Concurrency Anomaly Detection} \label{ssec:concurrency_anomaly_detection}

In practice, database transactions are often executed under reduced isolation levels, 
as there is an inherent trade-off between performance and isolation levels.
For example, both Oracle and Microsoft SQL Server default to read committed. 
Unfortunately, such weaker isolation levels can result in concurrency anomalies that
yield an interleaving of operations that could not arise in a serial execution of transactions. 
\codename provides a real-time breakdown of how many transactions may have experienced anomalies by running under reduced isolation levels (SI in particular).  It reports which transactions may have experienced anomalies and which transactions definitely did not.
\codename does this in a lightweight fashion that introduces minimal overhead, allowing developers to monitor
their production systems and tune the isolation levels on the fly.

Recall that a snapshot isolation transaction is assigned with one more timestamp \crts to validate its read set. 
According to the \codename protocol, a transaction having two equal timestamps, meaning the \crts is equal to the \cts, is serializable because all accesses occur at the same logical time. 
In the transaction validation phase, \codename applies this lightweight equality check to all SI transactions 
to detect transactions that may have observed concurrency anomalies (i.e., the ones with two different timestamps).

With the support of real-time anomaly detection, application developers can monitor how many anomalies arise with 
snapshot isolation in a timely manner and get insights to make better design decisions. 
For example, developers can switch to higher isolation levels when too many anomalies are detected 
or re-design application logic to eliminate anomalies in transactions running under reduced isolation levels. 

%% file: optimizations.tex
\section{Optimizations} \label{sec:optimizations}

In this section, we discuss two optimizations that further reduce network round trips. 

\subsection{Timestamp Synchronization} \label{ssec:timestamp_synchronization}

\codename can validate a previous read locally without communicating with the primary replica. 
We now introduce an optimization that boosts the system's performance by further reducing the frequency of remote validation.

\begin{figure}[!t]
\centering
\includegraphics[width=0.95\columnwidth]{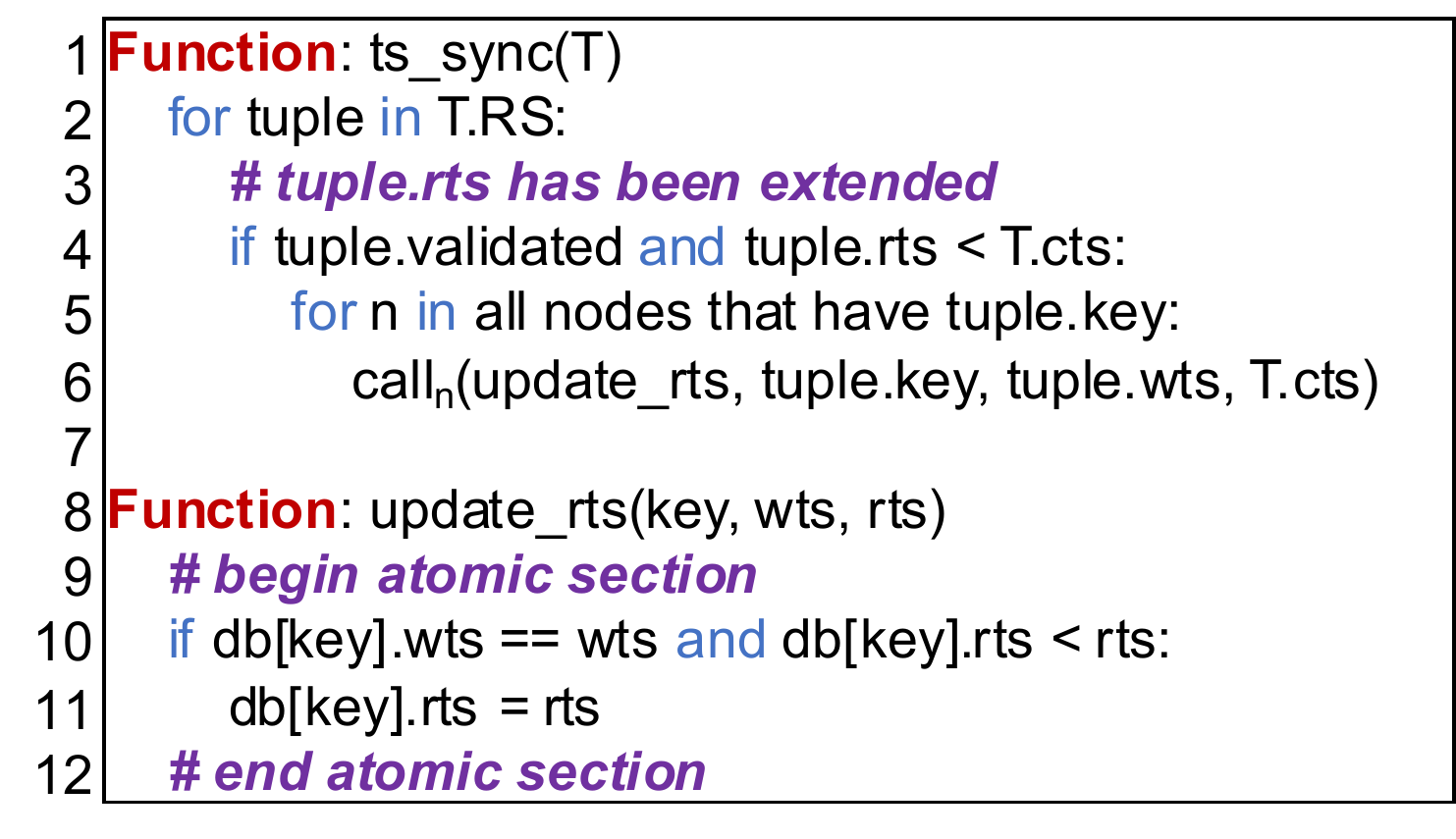}
\caption{Pseudocode to synchronize timestamps} \label{fig:ts_sync}
\end{figure}

The logical timestamps associated with each record on the primary replica are updated in two scenarios:
(1) a new value is written to a record when a transaction commits
(e.g., new timestamps with equal \wts and \rts are assigned), and
(2) the \rts is extended when a record is validated, as discussed in Section~\ref{ssec:transaction_validation}.
In the first scenario, a record's value and its associated timestamps are also updated on backup replicas.
However, the \rts of a record would not be updated on backup replicas in the second scenario by default.
As more and more transactions validate a record on the primary replica,
the gap between the \rts on the primary replica and backup replicas becomes larger.

Since the records on backup  replicas have stale and smaller \rts,
the record in a transaction's read set is less likely to be validated locally.
To address this problem, we apply an optimization we call {\it timestamp synchronization}.
The idea is to actively propagate the \rts from primary replicas to backup replicas.
As shown in Figure~\ref{fig:ts_sync}, the function \texttt{ts\_sync} is invoked
when a transaction commits or aborts (note that some records may be successfully validated even if a transaction aborts).
Since the value of a record is not included in this synchronization,
the system only extends the \rts when backup replicas have the same \wts,
 as shown in the function \texttt{update\_rts} (line 10 -- 11 of Figure~\ref{fig:ts_sync}). Note that timestamp synchronization happens asynchronously and thus does not increase the latency of a transaction.

\subsection{Parallel Locking and Validation} \label{ssec:parallel_locking_and_validation}

As we discussed in Section~\ref{ssec:transaction_validation}, if a transaction commits under serializability, the commit timestamp \cts must be larger than the \rts of each entry in the write set. 
Since a tuple's timestamp may be changed by other conflicting transactions, the latest \rts of each tuple is not available until it has been locked. 
In addition, a transaction must validate its read set with the \cts. 
In other words, a transaction cannot validate its read set until it has locked its write set. 

However, an SI transaction only requires that all reads are from a consistent snapshot. 
As we discussed in Section~\ref{sec:isolation}, an SI transaction has two timestamps (i.e., \crts and \cts). 
A transaction can calculate the \crts with the \wts of each entry in its read set.
With the \crts, the transaction is able to lock its write set and validate its read set in parallel.
Once all tuples in the write set have been locked, the transaction next calculates the \cts and commits if no conflicts exist.  

We now use an example to illustrate how \codename eliminates one network round trip  with the parallel locking and validation (PVL) optimization for SI transactions. 
 
\begin{exmp}
    Suppose an SI transaction reads tuple $x$ and $y$, and updates the value of tuple $x$ to $x + y$. The following operations are invoked: (1) Read $x$, (2) Read $y$, (3) Write $x$, and (4) Commit.
\end{exmp}

\begin{figure}[!t]
    \centering
    \includegraphics[width=0.9\columnwidth]{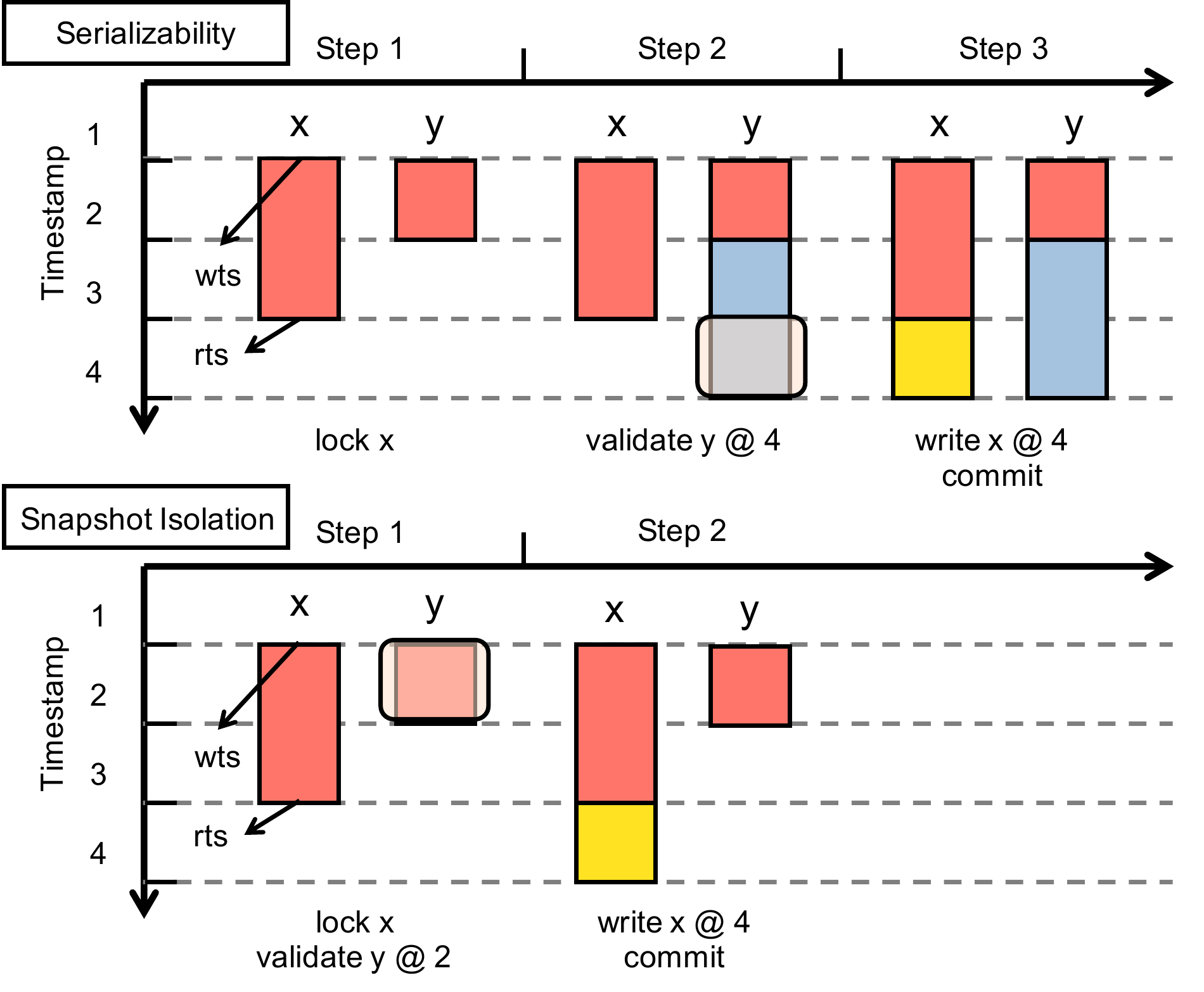}
    \caption{Illustrating parallel locking and validation} \label{fig:si_opt}
\end{figure}

We show a step-by-step diagram in Figure~\ref{fig:si_opt}, in which a tuple is shown as a vertical band. 
The start and end of a band indicate the tuple's \wts and \rts. Each step shows a different phase from a transaction's lifecycle. 
There are two steps to validate a serializable transaction and one step to validate an SI transaction. 
Both serializable and SI transactions have a step to commit in the end. 
A round trip communication may happen at the end of each step. For the ease of presentation, suppose there is no conflicting transaction that updates the timestamps of tuple $x$ and $y$.

The transaction reads tuple $x$:~\big[2,3\big] and $y$:~\big[2,2\big]. 
There are three steps in the transaction validation phase if the transaction commits under serializability (shown on the top of Figure~\ref{fig:si_opt}).

{\bf Step 1:} The transaction locks tuple $x$, since it's in the write set. According to the algorithm in Section~\ref{ssec:transaction_validation}, the \cts is 4, being the maximum of \wts in the read set and \rts $+ 1$ in the write set.

{\bf Step 2:} Tuple $y$ will be validated at timestamp 4. Since the \wts is not changed, it's not locked by other conflicting transactions, and the \rts of tuple $y$ can be extended to 4, the validation succeeds. 

{\bf Step 3:} The transaction updates tuple $x$ with a new value at the \cts and commits. 

As we discussed above, an SI transaction is able to lock the write set and validate the read set in a single step, as shown on the bottom of Figure~\ref{fig:si_opt}. 

{\bf Step 1:} The transaction first generates the timestamp for read validation. In this example, the maximum of \wts in the read set is 2. The transaction next uses this timestamp as the \crts to validate tuple $y$ and the validation succeeds. Meanwhile, the transaction also locks tuple $x$.

{\bf Step 2:} The transaction next generates the \cts, which is 4. At last, the transaction updates tuple $x$ and commits. 

%% file: discussion.tex

\section{Discussion} \label{sec:discussion}

As discussed in previous sections, \codename improves the performance of distributed OCC protocols through asynchronous replication and coordination reduction by using logical timestamps. \codename also allows transactions to read data from	backup replicas to reduce network messages. 

Besides \codename, some MVCC-based systems like Spanner~\cite{CorbettDEFFFGGHHHKKLLMMNQRRSSTWW12} and TAPIR~\cite{ZhangSSKP15} also allow transactions to read data from secondary replicas. 
As we will discuss in this section, however, MVCC-based systems may not be as effective as \codename in reducing coordination. 
Furthermore, we demonstrate that the technique in \codename can also be applied to an MVCC protocol as an improvement.

In an MVCC protocol, a transaction is assigned a unique commit timestamp at the beginning of its execution. 
The commit timestamp can be derived from either a synchronized clock (e.g., atomic clock~\cite{CorbettDEFFFGGHHHKKLLMMNQRRSSTWW12} or software-based solution~\cite{eidson2002ieee, mills1991internet}), or a centralized timestamp allocator~\cite{WuALXP17}. 
Each database record contains a version number which is the commit timestamp of the creating transaction. 
A transaction may read a record from a secondary replica, if its commit timestamp intersects the valid timestamp range for an old version of the record.

When a transaction is accessing the latest version of a record, however, an MVCC protocol may not be able to determine the consistency of the data based on the transaction's local information. 
Spanner~\cite{CorbettDEFFFGGHHHKKLLMMNQRRSSTWW12} solves this problem using atomic clocks, by letting the transaction wait until the uncertainty period has passed. This requires special hardware with atomic clocks which is very expensive. 
A more generic solution (e.g., TAPIR~\cite{ZhangSSKP15}) is to send a message to other replicas to check the consistency of the data. 
This introduces a least one round of network messages and therefore defeats the purpose of reading from replicas. 

The new idea that \codename introduces is the \rts of each record. 
With the \rts, a transaction knows that the data is guaranteed to be valid until that logical time and therefore is able to read the data without contacting other replicas. 
Note that the concept of \rts can also be applied to any MVCC protocols like Spanner~\cite{CorbettDEFFFGGHHHKKLLMMNQRRSSTWW12} or TAPIR~\cite{ZhangSSKP15} to reduce the number of coordination messages to see if the latest data version is read.

%% file: evaluation.tex
\section{Evaluation} \label{sec:evaluation}

In this section, we study the performance of \codename focusing on the following key questions: 

\begin{itemize}[noitemsep,nolistsep]
    \item How does \codename perform compared to other distributed concurrency control algorithms?
    \item How does network latency affect \codename?
    \item What's the performance of \codename with different numbers of replicas?
	\item How much performance gain can \codename achieve under snapshot isolation vs serializability?
	\item What fraction of transactions actually commit under serializability when they run under snapshot isolation?
	\item How effective is each optimization in \codename?
\end{itemize}

\subsection{Experimental Setup} \label{ssec:experimental_setup}

We run most of the experiments on a cluster of eight machines, each 
with 32 cores (four 8-core 2.13 GHz Intel(R) Xeon(R) E7-4830 CPUs) and 
256 GB of DRAM. Each machine runs 64-bit Ubuntu 12.04 with Linux 
kernel 3.2.0-23 and the servers are connected with a 10 GigE network.

In our experiments, we run 24 worker threads and 2 threads for network communication on each machine. 
Each worker thread has an integrated workload generator. 
Aborted transaction are re-executed with an exponential back-off strategy. 
All results are the average of ten runs. 

In Section~\ref{ssec:wide_area_network_experiment}, we run experiments 
on wide-area network setting using Amazon EC2 instances. 

\subsubsection{Workloads} \label{sssec:workloads}

To evaluate the performance of \codename, we ran a number of experiments 
using the following three popular benchmarks:

{\bf Retwis:} The Retwis benchmark is designed to model activities happened at Twitter~\cite{retwis}. 
There is a single table and each row is a key-value pair.  
We support two transactions, namely, (1) \texttt{PostTweet} and (2) \texttt{GetTimeline}.
A user can post a tweet to the social network via the \texttt{PostTweet} transaction. 
The \texttt{GetTimeline} transaction returns the latest tweets from a user and his/her followers.

{\bf YCSB:} The Yahoo! Cloud Serving Benchmark (YCSB) is a simple transactional workload.
It's designed to be a benchmark for facilitating performance
comparisons of database and key-value systems~\cite{CooperSTRS10}. 
There is a single table and each row has ten attributes. 
The primary key of the table is a 64-bit integer and each attribute has 10 random bytes.
Unless otherwise stated, a transaction consists of 4 operations in this benchmark.\footnote{YCSB+T~\cite{Dey13, DeyFNR14}, another extension to YCSB, wraps operations within transactions in a similar manner to model activities happened in a closed economy.}

{\bf TPC-C:} The TPC-C benchmark is a popular benchmark to evaluate OLTP databases~\cite{tpcc}. 
It models a warehouse-centric order processing application. 
We support the \texttt{NewOrder} transaction in this benchmark, which involves customers placing orders in their districts within a local warehouse. 
The local warehouse fulfills most orders but a small fraction of the orders involve products from remote warehouses.

In Retwis and YCSB, we set the number of records to 400K per partition and the number of partitions to 192, which equals to the total number of worker threads in the cluster.
To model different access patterns, we vary the skew factor and  the ratio of cross-partition transactions in our experiments (i.e., Section~\ref{ssec:performance_comparison}).  
In TPC-C, we set the number of warehouses to 192 as well. 
In all workloads, we set the number of replicas to 3, i.e., each partition has a primary partition and two secondary partitions, which are always hashed to three different nodes. 

\subsubsection{Distributed Concurrency Control Algorithms} \label{sssec:distributed_concurrency_control_algorithms}

By default, \codename is allowed to read from local secondary replicas. The timestamp synchronization optimization is also enabled, unless otherwise stated. 
We compared \codename with the following distributed concurrency control algorithms. 
To avoid an apples-to-oranges comparison, we implemented all algorithms in C++ in our framework. 
All systems are compiled using GCC 5.4.1 with the \texttt{-O2} option enabled.

{\bf \twopl:} This is a distributed concurrency control algorithm based on strict two-phase locking. Read locks and write locks are acquired as a worker runs a transaction. 
To avoid deadlock, the same \texttt{NO\_WAIT} policy is adopted as discussed in Section~\ref{sec:scar}. 
The worker updates all records and replicates the writes to replicas before releasing all acquired locks.

{\bf \occ:} This is a distributed optimistic concurrency control algorithm based on Silo's OCC protocol~\cite{TuZKLM13}. 
\occ assigns a transaction ID (TID) to each transaction 
when it commits based on the TID associated with each record in its read/write set. 
\occ also has the same three steps to validate a transaction as in \codename.
However, unlike \codename, it must validate all records of a transaction. This is achieved by comparing the data versions from the transaction's read set to the latest ones in the database on the primary replicas at commit time. 

{\bf \rc:} This is a reduced consistency protocol adapted from \occ. It supports {\it read committed} transactions by not validating a transaction's read set. 

By default, \occ and \rc are allowed to read from local secondary replicas. However, \rc does not need to use TIDs to validate its read set on the primary replicas, since every record it sees is guaranteed to come from committed transactions. \occ and \rc also use asynchronous writes and replication, and apply the same techniques as discussed in Section~\ref{ssec:asynchronous_write_and_replication} for consistency reasons.

A transaction runs under serializable isolation level in \codename, \twopl and \occ, and under read committed isolation level in \rc. 
\codename, \occ and \rc commit transactions with a group commit that happens every 10~ms.

\subsection{Performance Comparison} \label{ssec:performance_comparison}

We now study the performance of \codename and other algorithms using Retwis, YCSB and TPC-C workloads.

\begin{figure*}[!t]
    \centering
    \subfigure[50\% Cross-Partition]{\includegraphics[width=0.66\columnwidth]{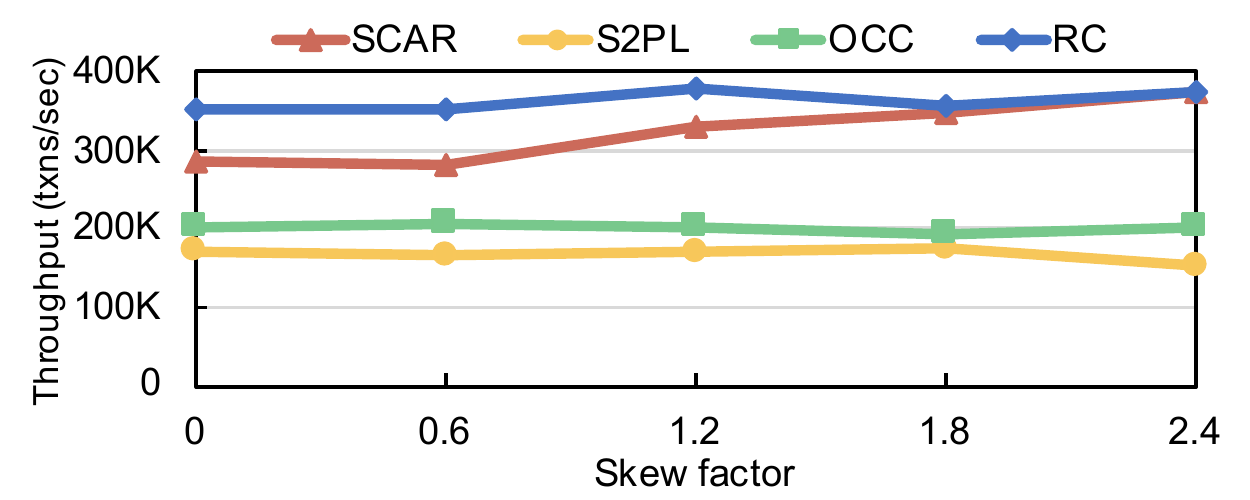}\label{fig:retwis_skew}}
    \subfigure[Skew factor = 1.2]{\includegraphics[width=0.66\columnwidth]{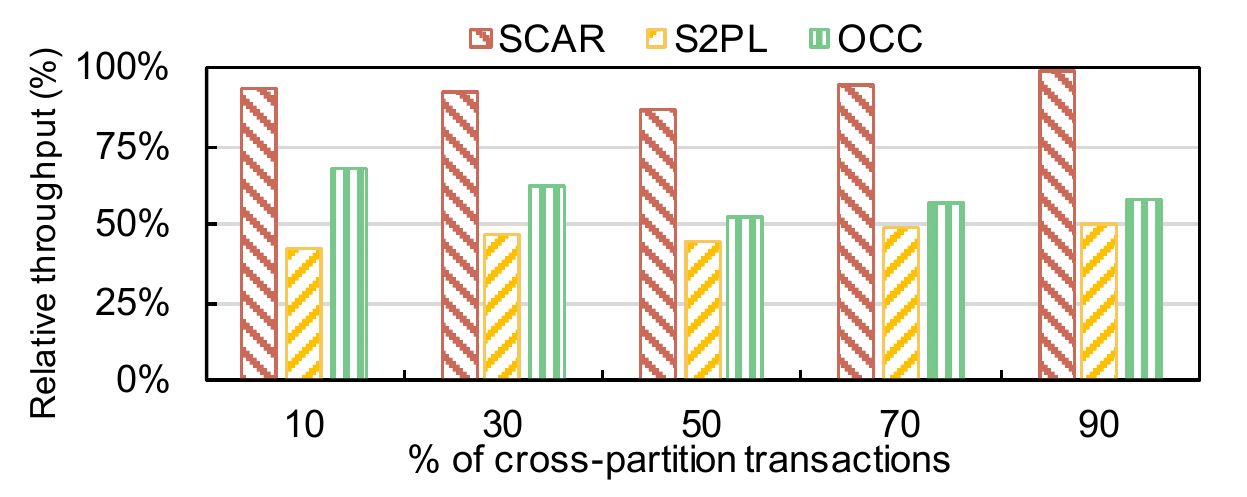}\label{fig:retwis_cross}} 
    \subfigure[\small 50\% Cross-Partition; Skew factor = 1.2]{\includegraphics[width=0.66\columnwidth]{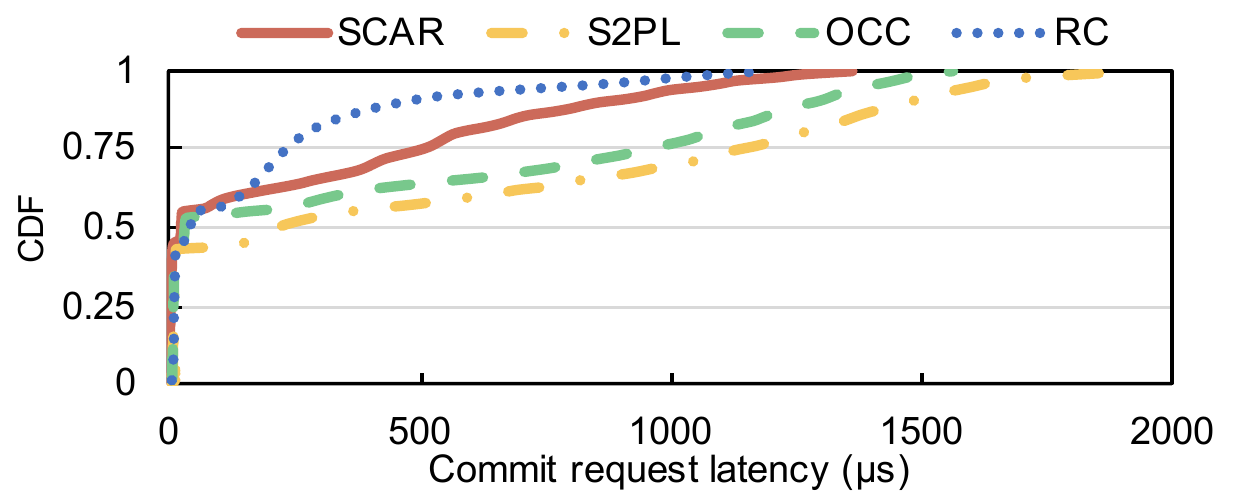}\label{fig:cdf_retwis}}
    \caption{Throughput and cumulative distribution of commit request latency of each approach on Retwis} \label{fig:retwis}
\end{figure*}

\begin{figure*}[!t]
    \centering
    \subfigure[50\% Cross-Partition]{\includegraphics[width=0.66\columnwidth]{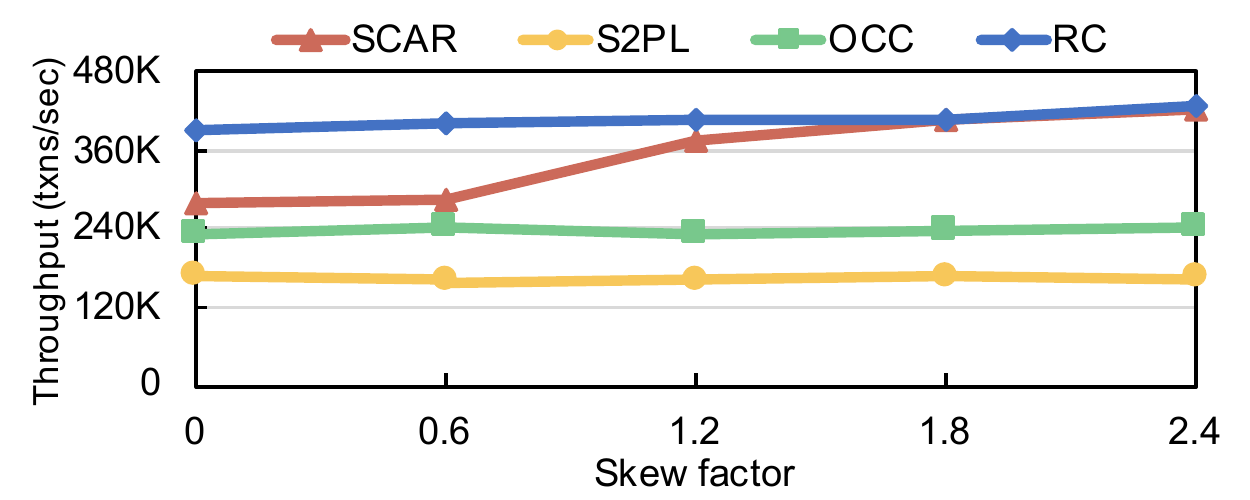}\label{fig:ycsb_skew}}
    \subfigure[Skew factor = 1.2]{\includegraphics[width=0.66\columnwidth]{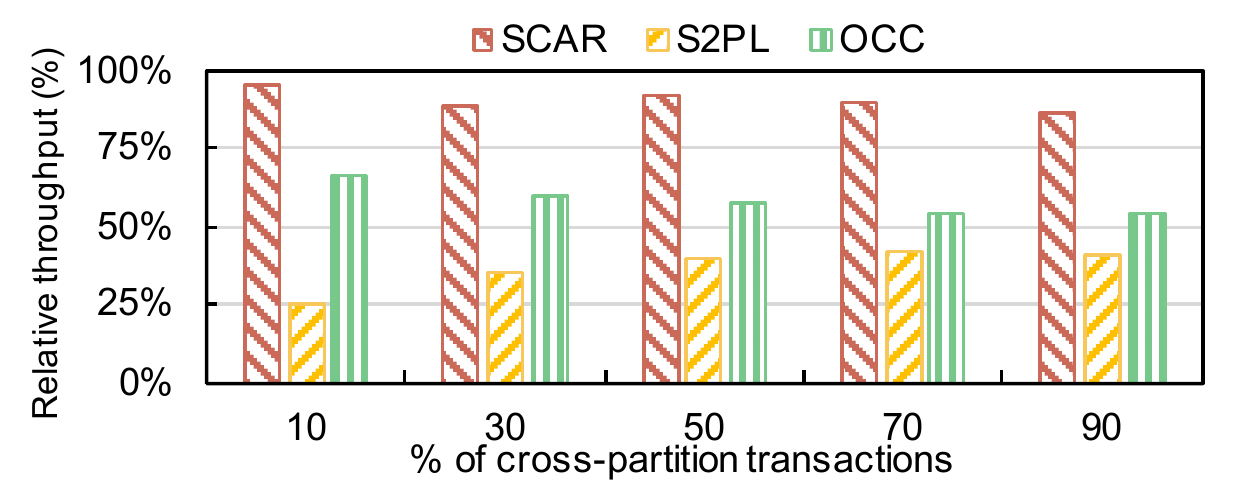}\label{fig:ycsb_cross}}
    \subfigure[\small 50\% Cross-Partition; Skew factor = 1.2]{\includegraphics[width=0.66\columnwidth]{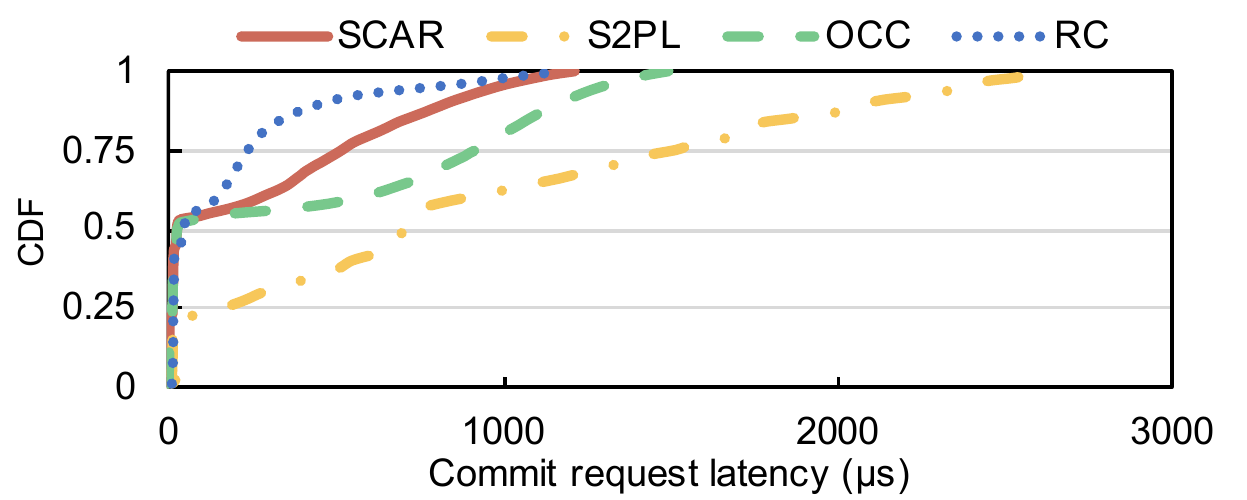}\label{fig:cdf_ycsb}}
    \caption{Throughput and cumulative distribution of commit request latency of each approach on YCSB} \label{fig:ycsb}
\end{figure*}

\subsubsection{Retwis Results} \label{sssec:retwis_results}
We first analyze the performance of all distributed concurrency control algorithms on the Retwis benchmark.
We run a workload mix of 80/20, i.e., the workload consists of 80\% of the \texttt{GetTimeline} transaction and 20\% of the \texttt{PostTweet} transaction.

The \texttt{GetTimeline} transaction has $n$ read operations that load tweets 
from a given user and his/her followers, where $n$ is chosen at random from 1 to 10. 
There are 3 update operations and 2 write operations in the \texttt{PostTweet} transaction.
In social network, some popular tweets are read by a lot more people. 
To model this, each operation in the \texttt{PostTweet} transaction follows a uniform distribution, but each operation in the \texttt{GetTimeline} transaction follows a Zipfian distribution~\cite{GraySEBW94} with a skew factor.

Figure~\ref{fig:retwis_skew} shows the results with a varying skew factor and 50\% cross-partition transactions.
\twopl has consistently lower throughput than other algorithms, since it always reads from primary replicas and applies writes synchronously.
\rc has the highest throughput, since it avoids much coordination by running transactions under read committed. 
When the skew factor is 0, i.e., each access follows a uniform distribution, \codename has 41\% higher throughput than \occ. 
This is because \occ has to validate every records it reads. In contrast, \codename can locally validate some records as discussed in Section~\ref{sec:scar}. 
As we increase the skew factor, the \rts of each record is more likely to be valid at a transaction's commit timestamp in \codename. For this reason, with a skew factor being 2.4, \codename has 1.9x higher throughput than \occ and achieves similar throughput to \rc.

We also ran Retwis with a fixed skew factor 1.2 and a varying ratio of cross-partition transactions. 
Since the throughput of each approach decreases significantly, 
we report each approach's relative throughput to \rc in Figure~\ref{fig:retwis_cross} for the purpose of better visualization. 
Overall, \codename achieves up to 70\% higher throughput than \occ.

We now study the commit request latency of each approach.
The commit request latency measures how long it takes a transaction to release its write locks since the beginning of execution~\cite{AthanassoulisJAS09}.
The overall execution latency is not reported due to the fact that all approaches except \twopl use a group commit. We report the cumulative distribution function (CDF) of the commit request latency in each approach in Figure~\ref{fig:cdf_retwis}. 
We fixed skew factor to 1.2 and the ratio of cross-partition transactions to 50\%. 
In Figure~\ref{fig:cdf_retwis}, we can observe that \codename has consistently lower commit request latency than \occ and \twopl. 
The commit request latency of \rc is the lowest, since it runs transactions under read committed and avoids more coordination than others.

\subsubsection{YCSB Results} \label{sssec:ycsb_results}

We next study the performance of \codename versus the other algorithms on the YCSB benchmark. 
As in Retwis, each read operation follows a Zipfian distribution~\cite{GraySEBW94} and each update operation follows a uniform distribution.
We run a workload mix of 80/20, i.e., each operation in a transaction has an 80\% probability of being a read operation and a 20\% probability of being an update operation.

Figure~\ref{fig:ycsb_skew} shows the throughput of each approach with a varying skew factor and 50\% cross-partition transactions. 
We observe a similar result as for Retwis. For example, the throughput of \codename goes up as we increase the skew factor. 
The throughput of other approaches is not sensitive to the skew factor. When the skew factor is 2.4, \codename has 75\% higher throughput than \occ and achieves similar throughput to \rc. 
Figure~\ref{fig:ycsb_cross} shows the relative throughput of each approach to \rc with a varying ratio of cross-partition transactions. We fix the skew factor to 1.2 as well.
Overall, \codename has up to 65\% higher throughput than \occ. 

Figure~\ref{fig:cdf_ycsb} shows the CDF of the commit request latency on YCSB. 
We fixed skew factor to 1.2 and the ratio of cross-partition transactions to 50\%. 
As in Retwis, we can observe that the gap is smaller between \codename and \rc than \occ and \twopl.

\subsubsection{TPC-C Results} \label{sssec:tpcc_results}

\begin{figure*}[!t]
    \hspace{-0.015\linewidth}
    \begin{minipage}[!t]{0.48\linewidth}
    \centering
    \includegraphics[width=0.85\columnwidth]{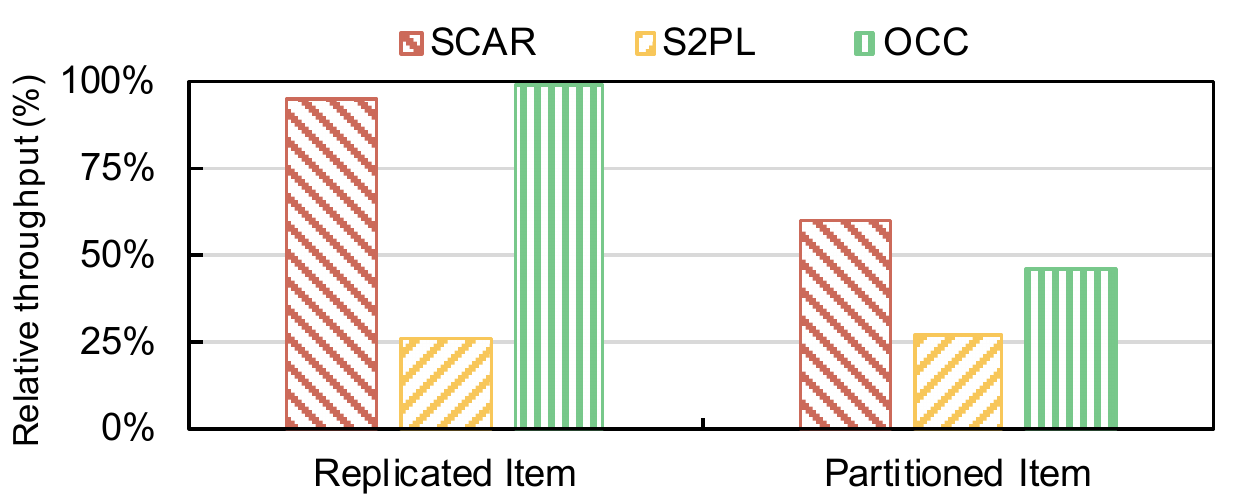}   

    \caption{Throughput of each approach on TPC-C} \label{fig:tpcc}   
    \end{minipage}
    \hspace{0.03\linewidth}
    \begin{minipage}[!t]{0.48\linewidth}
    \centering
    \includegraphics[width=0.85\columnwidth]{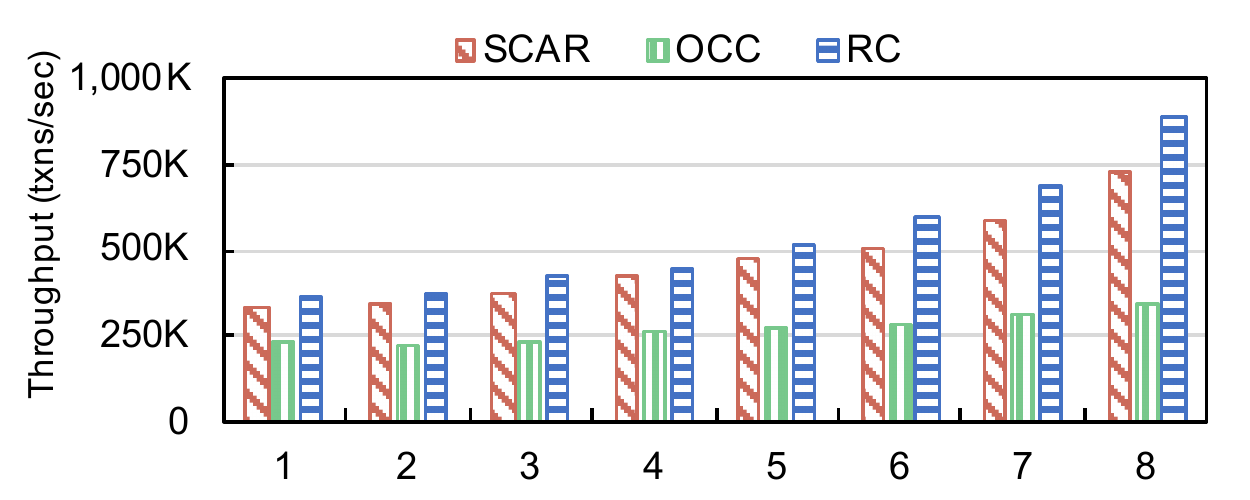}      

    \caption{Results with a varying number of replicas}\label{fig:replication_factor}
    \end{minipage}

\end{figure*}

At last, we study the performance of each approach on the TPC-C benchmark. We ran a workload with the \texttt{NewOrder} transaction only. 
10\% of transactions are fulfilled by a remote warehouse, meaning that they are cross-partition transactions. 
We consider two variations of TPC-C in this experiment: (1) \textit{Replicated Item}: the \texttt{Item} table is replicated on each node and considered to be a read only index; (2) \textit{Partitioned Item}: the \texttt{Item} table is partitioned across all nodes the same as all other tables. 
In the second variation, more transactions become cross-partition transactions. This is because every \texttt{NewOrder} transaction has 5~$\sim$~15 reads from the \texttt{Item} table. If no local replica is available, these reads are remote. 

For the purpose of better visualization, we report each approach's relative throughput to RC as we did in Section~\ref{ssec:performance_comparison}.
We show the result of \textit{Replicated Item} in the left side of Figure~\ref{fig:tpcc}. Since the \texttt{NewOrder} transaction is a write-intensive transaction, i.e., almost every read comes with an update\footnote{The reads from the \texttt{Warehouse} and the \texttt{Customer} table are always local.}, this benchmark gives no benefits to \codename and makes it slightly slower than \occ due to more messages being sent (e.g., messages to synchronize timestamps as discussed in Section~\ref{ssec:timestamp_synchronization}).
The result of \textit{Partitioned Item} is shown in the right side of Figure~\ref{fig:tpcc}. With more reads from the \texttt{Item} table being remote, \codename achieves 32\% higher throughput than \occ because of coordination reduction. 
\rc has the highest throughput, since it never validates remote reads.

In summary, \codename is able to achieve higher throughput than \occ and \twopl by reducing coordination. 
In the case that there exists high access skew in read operations, its performance is even close to running transactions under reduced isolation levels (e.g., read committed).

\subsection{Wide-Area Network Experiment} \label{ssec:wide_area_network_experiment}

\begin{table}
  \centering
  \includegraphics[width=0.85\columnwidth]{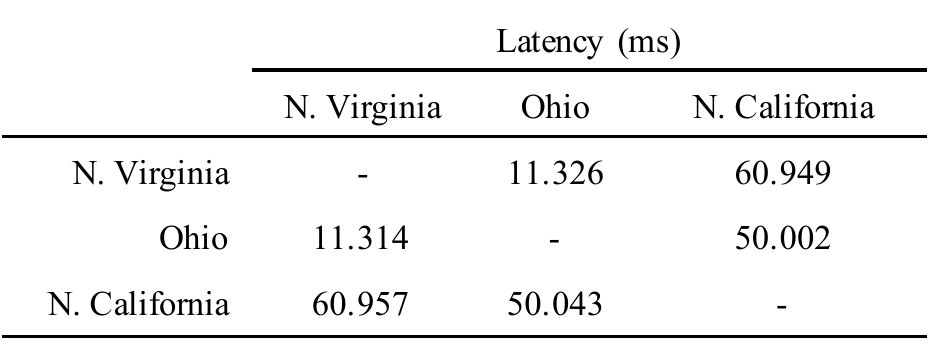}
  \caption{Round trip times between EC2 nodes} \label{tbl:ec2_latency}
\end{table}

\begin{figure}[!t]
  \centering
  \includegraphics[width=0.85\columnwidth]{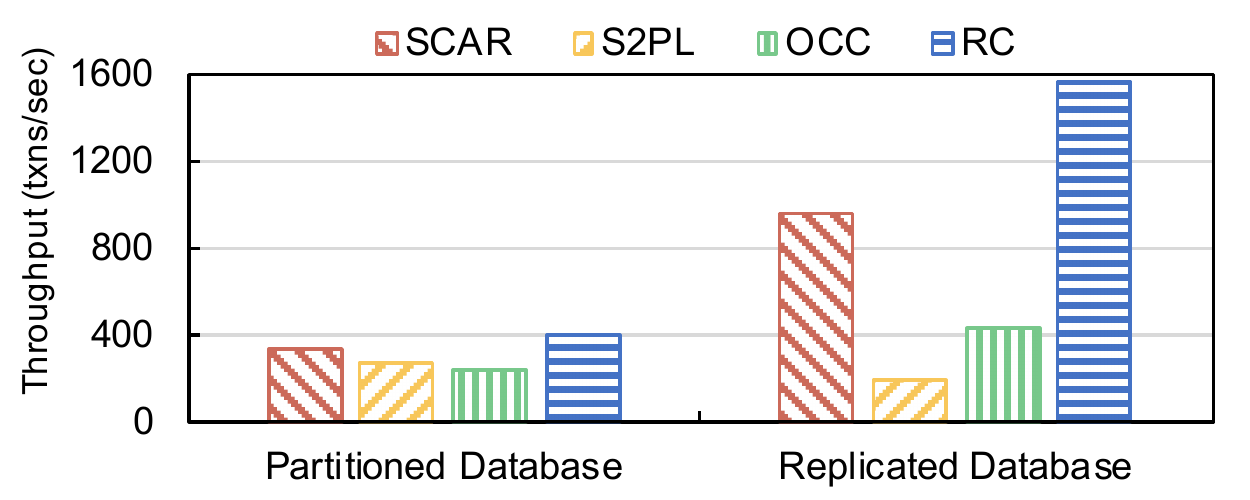} 
  \caption{Results in wide-area network setting} \label{fig:ec2}
\end{figure}

In this section, we study how \codename performs compared to other approaches in the wide-area network (WAN) setting. 
For users concerned with very high availability, wide-area replication is important because it allows the database to survive the failure of a whole data center (e.g., due to a power outage or a natural disaster).
We use three m5.2xlarge nodes running on Amazon EC2~\cite{ec2}. The three nodes are in North Virginia, Ohio, and North California respectively and each node has 8 virtual CPUs. 
The round trip times between any two nodes are shown in Table~\ref{tbl:ec2_latency}. 
We consider two variations in this experiment: (1) \textit{Partitioned Database}: the database is partitioned across three area zones and no replication is used; (2) \textit{Replicated Database}: each partition of the database is fully replicated in all area zones, meaning each one has a primary partition and two backup partitions. The primary partition is randomly chosen from 3 nodes.

In this experiment, we run 6 worker threads and 2 threads for network communication on each node.
The group commit frequency is set to once per second.  The same YCSB workload from Section~\ref{sssec:ycsb_results} is used, with the skew factor being 1.2 and 50\% cross-partition transactions.

As we can observe from Figure~\ref{fig:ec2}, \rc has the highest throughput in both scenarios, since less coordination is required. 
In addition, all approaches except \twopl have higher throughput when the database is replicated across area zones. 
The reasons are twofold: (1) \twopl uses synchronous write and replication. More replicas lead to higher latency and lower throughput; 
(2) all other protocols are optimistic and are able to take advantage of local replicas to reduce network communication. In \textit{Partitioned Database}, \codename achieves 26\% higher throughput than \twopl and 41\% higher throughput than \occ. 
In \textit{Replicated Database}, the performance of \codename and \rc is significantly improved. For example, in the experiment running in the local area network setting, \codename achieves only 62\% higher throughput than \occ (i.e., Figure~\ref{fig:ycsb_skew}). With wide-area network, \codename's performance improvement over \occ is 2.3x.

Overall, the performance advantage of \codename over other approaches is even more significant in the WAN setting.

\subsection{Effect of Different Numbers of Replicas}

We now study how \codename performs with different numbers of replicas. In this experiment, we report the results on the YCSB benchmark in Figure~\ref{fig:replication_factor}, with the skew factor being 1.2 and 50\% cross-partition transactions. 
We varied the number of replicas from 1 (only the primary replica exists) to 8 (one primary replica plus seven backup replicas) for each partition. 

Since the workload is read-intensive, each approach is expected to have higher throughput when more replicas are available. This is because  writes and replication are not a bottleneck in this workload, and more reads requests are served locally.
For example, \rc almost achieves 2.4x higher throughput when 8 replicas are available. 
In contrast, \occ achieves only 52\% higher throughput, since many reads still need validation. 
\codename reduces coordination through logical timestamps and achieves 2.2x higher throughput.

In summary, reading from local replicas effectively boosts a system's performance. 
By default, each partition in \codename has three replicas.

\begin{figure*}[!t]
    \hspace{-0.015\linewidth}
    \begin{minipage}[!t]{0.48\linewidth}
    \centering
    \includegraphics[width=0.85\columnwidth]{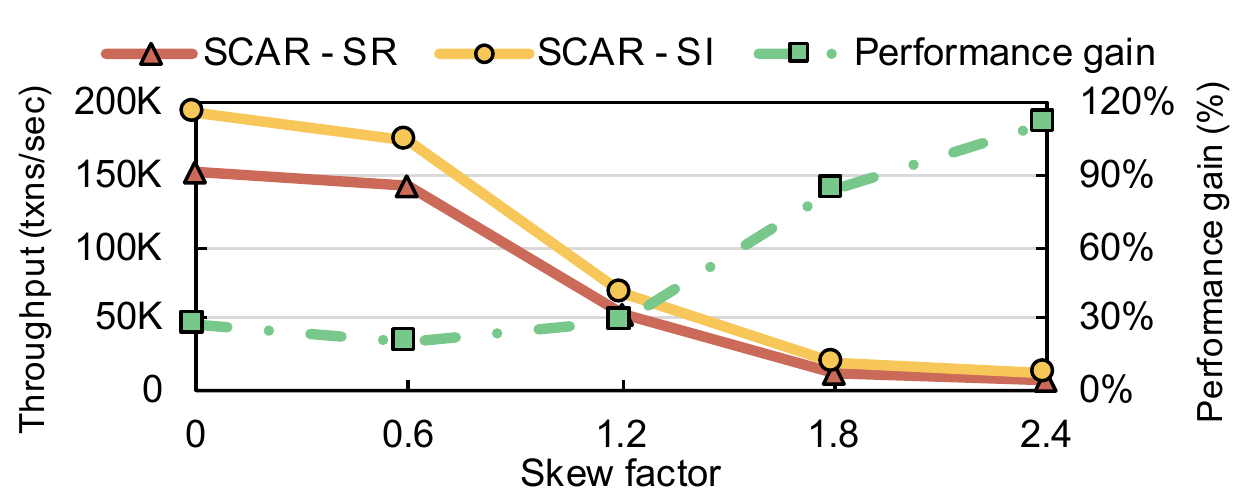}   

    \caption{Serializability (SR) vs. Snapshot Isolation (SI); Dotted line shows the performance gain of SI over SR} \label{fig:sr_vs_si}   
    \end{minipage}
    \hspace{0.03\linewidth}
    \begin{minipage}[!t]{0.48\linewidth}
    \centering
    \includegraphics[width=0.85\columnwidth]{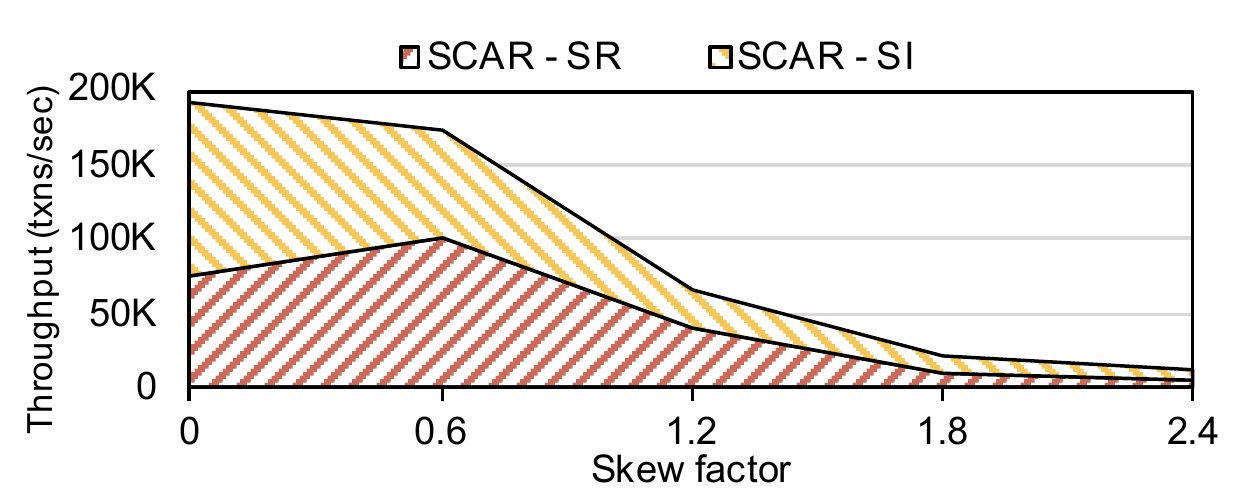}      

    \caption{Percentage of serializable (SR) transactions running under snapshot isolation (SI); Red area shows the throughput of SR transactions}\label{fig:si_in_sr}
    \end{minipage}

\end{figure*}

\subsection{Serializability vs. Snapshot Isolation} \label{ssec:serializability_vs_snapshot_isolation}

We next study how much performance gain \codename is able to achieve when it runs transactions under snapshot isolation (SI) versus serializability. 
Our intuition is that the system should commit more transactions per second under SI. 
The reasons behind are twofold. First, SI transactions does not detect read/write conflicts, which introduces a lower abort rate in a highly contended workload. 
Second, \codename running under SI can lock a transaction's write set and validate its read set in a single round trip as discussed in Section~\ref{ssec:parallel_locking_and_validation}.

In this experiment, we only report the results on the YCSB benchmark. 
This is because, in the Retwis benchmark, the \texttt{GetTimeline} transaction is a read-only transaction and these two isolation levels are the same to it, and TPC-C is a write-intensive benchmark. 
To increase read/write conflicts, we set the number of operations to 8 on YCSB and all operations follow a Zipfian distribution with a skew factor.

We vary the skew factor from 0 to 2.4. The ratio of cross-partition transactions is 50\% and the workload mix is 80/20.

We report the results in Figure~\ref{fig:sr_vs_si}. 
Two solid lines show the throughput of \codename running under different isolation levels. 
The dashed line shows the performance gain of running under SI over serializability.
As there is more contention in the workload, the performance of the system running under two different isolation levels goes down. 
This is because the system has a higher abort rate when the workload becomes more contended. 
As expected, \codename running under SI has a larger performance gain when the skew factor increases. 
For example, the performance gain goes up from 27\% to 111\%, as the skew factor goes up from 0.6 to 2.4. 

\subsection{Concurrency Anomaly Detection} \label{ssec:exp:concurrency_anomaly_detection}

There is an inherent trade-off between throughput and isolation levels. 
A system runs more transactions under SI than serializability, 
but concurrency anomalies may arise. 
As we discussed in Section~\ref{ssec:concurrency_anomaly_detection}, \codename is able to give a real-time breakdown
if a transaction commits under serializability when it started as a snapshot isolation transaction.

We used the same workload as in Section~\ref{ssec:serializability_vs_snapshot_isolation} 
and report the throughput of \codename running transactions under SI as well as serializability in Figure~\ref{fig:si_in_sr}.
As the workload has more contention, the percentage goes down. For example,
When the skew factor is 0.6, 58\% of SI transactions commit under serializability.   
The percentage goes down to 40\% as we increase the skew factor to 2.4. 
This is because fewer transactions are able to commit under a higher isolation level in a highly contended workload.

In summary, there are a large number of transactions that actually commit under serializability when they started as SI transactions. 
We believe application developers can achieve higher performance with \codename running under SI and monitor how many concurrency anomalies arise at the same time, with an eye towards switching to serializable mode if too many transactions are not achieving serializability.

\begin{figure*}[!t]
    \centering
    \subfigure[Retwis]{\includegraphics[width=0.66\columnwidth]{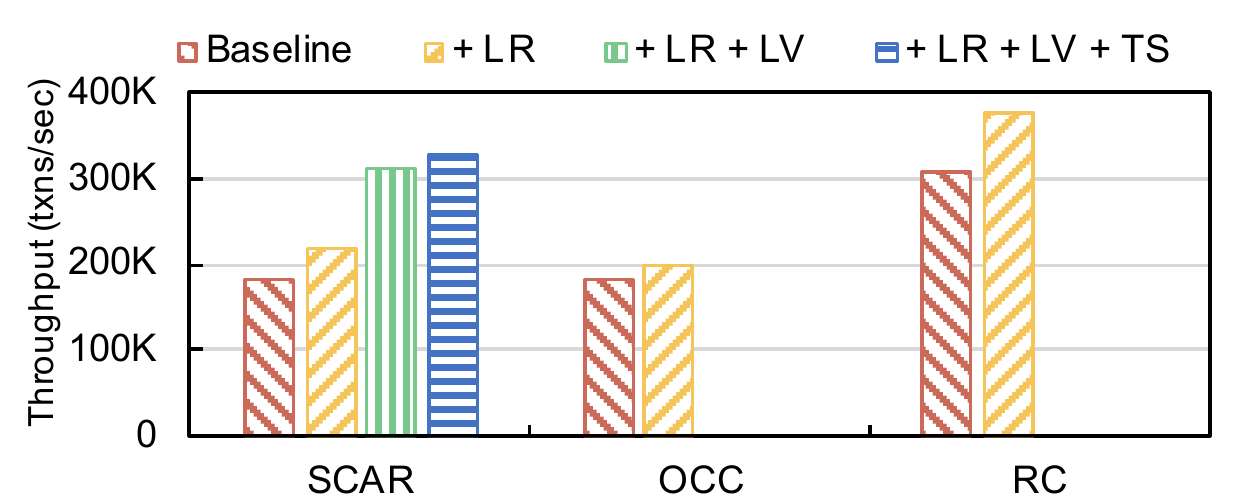}\label{fig:opt_retwis}}
    \subfigure[YCSB]{\includegraphics[width=0.66\columnwidth]{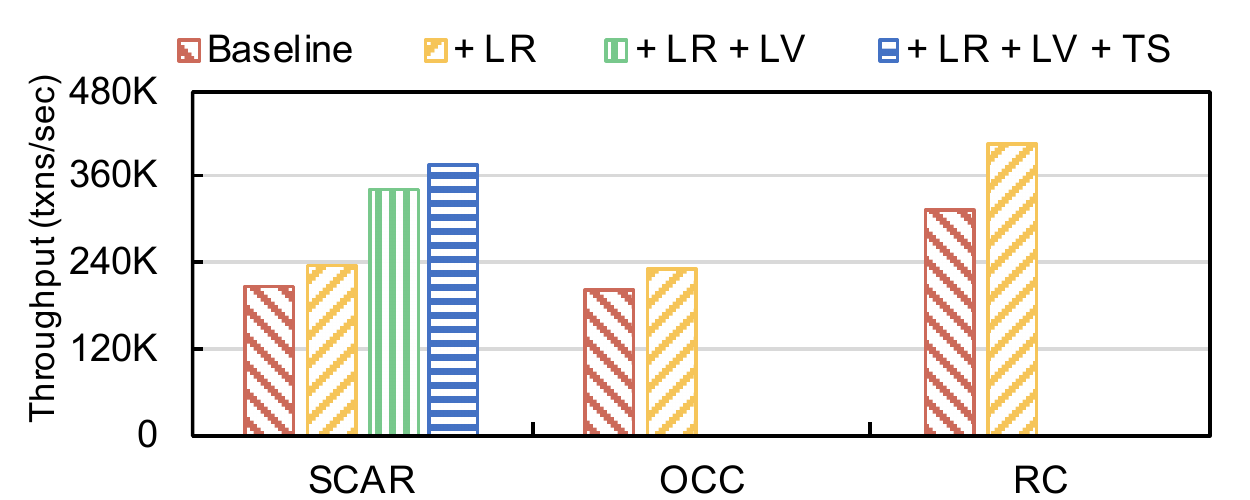}\label{fig:opt_ycsb}}
    \subfigure[YCSB]{\includegraphics[width=0.66\columnwidth]{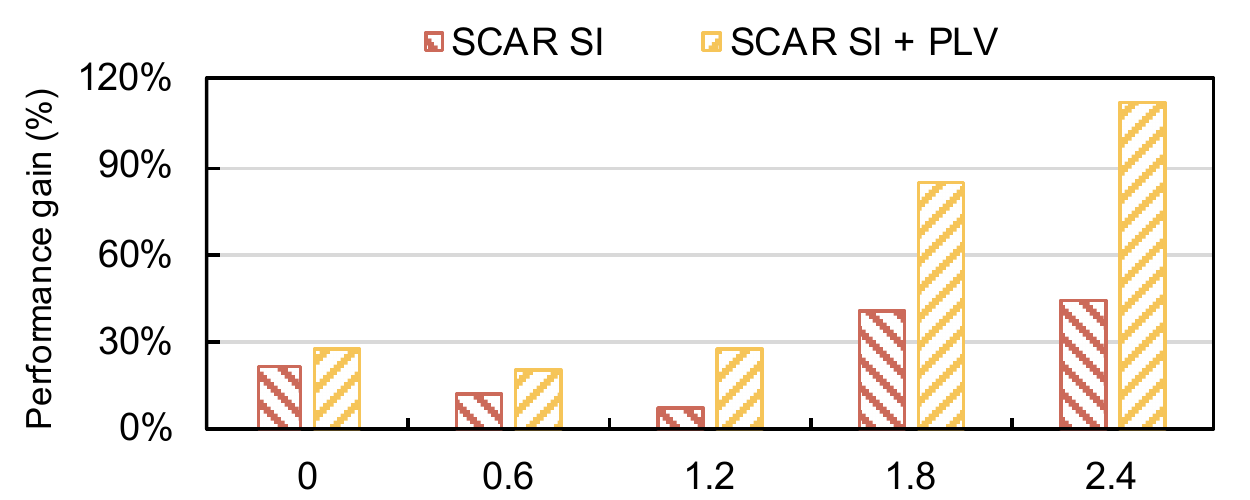}\label{fig:opt_plv}}
    \caption{Factor analysis for \codename, \occ and \rc; 50\% Cross-Partition; Skew factor = 1.2; LR=local read, LV=local read validation;  TS=timestamp synchronization; PLV=parallel locking and validation}\label{fig:opt}
\end{figure*}

\subsection{Factor Analysis} \label{ssec:factor_analysis}

We now study the effectiveness of each optimization technique in more detail through a factor analysis. 

\subsubsection{LR, LV and TS Results}\label{sssec:lr_lv_and_ts_results}

In the baseline implementation, \codename is only allowed to read from primary replicas and
validation messages are always sent to primary replicas to better study the effectiveness of local read validation.
The timestamp synchronization, as discussed in Section~\ref{ssec:timestamp_synchronization}, is also disabled. 
Similarly, the baseline implementations of \occ and \rc always read from primary replicas as well. 
We don't show the results on \twopl in this experiment, since it is not clear to us how to apply these optimization techniques to it.

We introduce one technique at a time to the baseline implementation and 
show the results in Figure~\ref{fig:opt_retwis} and Figure~\ref{fig:opt_ycsb}.
+LR refers to the technique that allows a transaction to read from local secondary replicas. 
+LR+LV refers to adding the local read validation on top of \codename when the first technique is enabled. 
A record in a transaction's read set can be locally validated if 
its \rts ends after the transaction's commit timestamp. 
Finally, +LR+LV+TS refers to adding the timestamp synchronization optimization 
with the first two techniques enabled. 

We first ran the same Retwis workload as we did in Figure~\ref{fig:retwis_skew} 
with a skew factor being 1.2. The results are shown in Figure~\ref{fig:opt_retwis}. 
The +LR technique enables \codename to have 20\% performance gain. 
Similarly, the +LR technique also helps \occ and \rc achieve 10\% and 23\% performance gain respectively.
As we further add the local read validation technique to \codename (shown as +LR+LV), the performance gain goes up to 1.7x compared to the baseline implementation. 
When the first two techniques are used with the timestamp synchronization (shown as +LR+LV+TS), 
\codename is able to achieve 1.8x higher throughput in total than the baseline implementation.

Similar results are also observed on the YCSB workload as shown in Figure~\ref{fig:opt_ycsb}.
The workload is the same as in Figure~\ref{fig:ycsb_skew} with a skew factor being 1.2. 
The +LR technique helps \occ and \rc achieve 15\% and 29\% performance gain respectively. 
\codename is able to achieve 1.8x higher throughput with all three techniques enabled (shown as +LR+LV+TS). 

\subsubsection{PLV Results} 
We now study the effectiveness of the parallel locking and validation (PLV) optimization. 
This technique potentially reduces one network round trip for SI transactions.

As we discussed in Section~\ref{ssec:parallel_locking_and_validation}, the +PVL technique can only be applied to SI transactions. 
For this reason, we ran the workload from Section~\ref{ssec:serializability_vs_snapshot_isolation}.
In Figure~\ref{fig:opt_plv}, the baseline implementation, which is shown as \codename~SI,
refers to \codename running under SI with all three techniques from Section~\ref{sssec:lr_lv_and_ts_results} enabled.
\codename~SI~+~PVL refers to adding the parallel locking and validation optimization on top of \codename~SI. The results are reported as the performance gain of running under SI over serializability.

When each operation follows a uniform distribution, the +PLV technique enables \codename~SI to have 5\% more performance gain. 
This is because one network round trip is eliminated.
As we add more contention to the workload, the additional performance gain goes up to 67\%. 
This is because the +PLV technique can effectively reduce the time that an SI transaction holds locks as well, allowing the system to have a lower abort rate and higher throughput.

%% file: related_work.tex
\section{Related Work} \label{sec:related_work}

The design of \codename is inspired by many pieces of related work, 
inducing transaction processing, strong consistency with replication and snapshot isolation systems.

{\it Transaction Processing.} The seminal survey by Bernstein et al.~\cite{BernsteinG81} summarizes classic distributed concurrency control protocols, with the exception of optimistic concurrency control~\cite{KungR81}. 
As in-memory databases becoming more popular, there has been a resurgent interest in transaction processing in both multicore processors~\cite{KimWJP16, LimKA17, TuZKLM13, WuALXP17, YuPSD16} and distributed systems~\cite{CowlingL12, HardingAPS17, MahmoudANAA14, MuCZLL14}. None of these protocols, however, provide high availability via replication. 

Query Fresh~\cite{WangJP17} uses an append-only storage architecture to 
make backup nodes of  a hot standby system not to block the primary node. 
\codename can use the same technique to further decrease the overhead of replication. 
Obladi~\cite{Crooks0CHAA18} reduces bandwidth cost and increases system throughput by delaying updates within epochs. 
Similarly, \codename uses asynchronous writes and replication to increase system throughput with epochs as well.

{\it Strong Consistency with Replication.} High availability is typically implemented via replication. Paxos~\cite{Lamport01} is a popular solution to coordinate the concurrent reads and writes to different copies of data while providing consistency. Spanner~\cite{CorbettDEFFFGGHHHKKLLMMNQRRSSTWW12} is a Paxos-based transaction processing system based on a two-phase locking protocol. Each read request goes to the master replica which can be a remote node. Each master replica initiates a Paxos protocol to synchronize with backup replicas. The protocol incurs multiple round-trip messages for data accesses and replication coordination. 
MDCC~\cite{KraskaPFMF13} is an OCC protocol that exploits generalized Paxos~\cite{Leslie05} to reduce the coordination overhead where a transaction can commit with a single message round trip in the normal operation. 
Ganymed~\cite{PlattnerA04, PlattnerAO08} runs update transactions on a single node and propagates writes of committed transactions to a  potentially unlimited number of read-only replicas.
TAPIR~\cite{ZhangSSKP15} eliminates the overhead of Paxos by allowing inconsistency in the storage system and building consistent transactions using inconsistent replication. Similar to \codename, TAPIR uses an optimistic protocol to validate transactions. The behavior of TAPIR is similar to the \occ configuration in Section~\ref{sec:evaluation}, and suffers its same limitations. While the systems above are different from the primary-backup design in \codename, the use of logical timestamps to reduce coordination among replicas is applicable to these systems as well. We leave the exploration of this to future work.

By maintaining multiple data versions, TxCache~\cite{PortsCZML10} ensures that a transaction always reads from a consistent snapshot regardless of whether each read operation comes from the database or the cache. 
In \codename, reads are from a consistent snapshot as long as they can be validated at a given logical time. 
Warranties~\cite{LiuMAGM14} reduces coordination on read validation by maintaining time-based leases to popular records, but writes have to be delayed. \codename reduces coordination without penalizing writes. This is because writes instantly make the read validity timestamps on old records expired.

{\it Isolation Levels.} Berenson et al.~\cite{BerensonBGMOO95} provides an excellent explanation of commonly used isolation levels in a database. 
Binning et al.~\cite{BinnigHFKLM14} show how only distributed SI transactions pay the cost of coordination. 
In \codename, cross-node coordination for local transactions is not necessary as well due to the use of logical timestamps. 
Serial safety net (SSN)~\cite{WangJFP17} is able to make any concurrency control protocol to support serializability by detecting dependency cycles. 
The same technique can be applied to \codename as well. Furthermore, 
\codename is able to monitor concurrency anomalies from SI transactions through a simple equality check.  

Due to the overhead of implementing strong isolation, many systems use weaker isolation levels instead (e.g., PSI~\cite{SovranPAL11}, causal consistency~\cite{MehdiLCABL17}, eventual consistency~\cite{TerryTPDSH95}, or no consistency~\cite{RechtRWN11}).
Lower isolation levels trade programmability for performance and scalability. In this paper, we focus on serializability and snapshot isolation, which are the gold standard for transactional applications and the default isolation levels in all major relational systems.

%% file: conclusion.tex
\section{Conclusion} \label{sec:conclusion}

In this paper, we presented \codename, a new distributed and replicated in-memory database. 
It allows transactions to read from secondary replicas and enforces strong consistency.
The writes of transactions are asynchronously replicated to secondary replicas and applied in any order. 
By serializing transactions in a logical-time order, the system 
is able to enforce strong consistency without expensive coordination. 
Our results on three popular benchmarks show that the system outperforms conventional designs by up to a factor of 2. 
In workloads with high contention, we also demonstrated that higher throughput can be achieved 
by running transactions under reduced isolation levels and 
the system can effectively monitor concurrency anomalies in real time. 